\documentclass[11pt]{article}

\usepackage{amsmath}
\usepackage{amsfonts}
\usepackage{amsthm}
\usepackage[english]{babel} 
\usepackage{dsfont}
\usepackage{amssymb}

\usepackage{graphicx}        
\usepackage{cite}   
\usepackage{booktabs}
\usepackage{caption}
\usepackage{subcaption}
\usepackage{epstopdf}
\usepackage{epsfig}
\usepackage[table]{xcolor}
\usepackage{subfloat}
\usepackage{float}
\usepackage{array}
\usepackage{mathtools}
\usepackage{bm}
\usepackage{mathrsfs}
\usepackage{hyperref}

\newcommand\B{\mathscr B}
\newcommand\C{\mathscr C}
\newcommand\F{\mathscr F}
\newcommand\E{\mathscr E}
\newcommand\R{\mathbb R}

\newcommand\de{\mathrm d}
\newcommand\e{\mathrm e}
\newcommand\der[2]{\frac{\de #1}{\de #2}}                 % standard derivatives
\newcommand\pd[2]{\frac{\partial #1}{\partial #2}}         % partial derivatives
\renewcommand\*{_\ast}

\let\contenutoin=\subseteq
\let\epsilon=\varepsilon
\let\ft=\hat                 % Fourier transform

\newtheorem{theorem}{Theorem}

\DeclarePairedDelimiter{\ins}{\lbrace}{\rbrace}                            % set
\DeclarePairedDelimiter{\abs}{\lvert}{\rvert}                   % absolute value
\DeclarePairedDelimiter{\expvalue}{\langle}{\rangle}            % expected value
\DeclarePairedDelimiter{\norminf}{\lVert}{\rVert_\infty}         % infinity norm

\title{Spreading of fake news, competence, and learning: kinetic modeling and numerical approximation}

\author{Jonathan Franceschi\thanks{Department of Mathematics and Computer 
Science, University of Ferrara, Via Machiavelli 30, 44121 Ferrara, Italy 
(jonathan.franceschi@edu.unife.it)} \and Lorenzo Pareschi\thanks{Department of 
Mathematics and Computer Science, University of Ferrara, Via Machiavelli 30, 
44121 Ferrara, Italy (lorenzo.pareschi@unife.it)}\ \thanks{Center for 
Modeling, Computing and Statistic (CMCS), University of Ferrara, Via Muratori 
9, 44121 Ferrara, Italy}}

\begin{document}
\maketitle

\begin{abstract}
The rise of social networks as the primary means of communication in almost 
every country in the world has simultaneously triggered an increase in the 
amount of fake news circulating online. This fact became particularly evident 
during the 2016 U.S. political elections and even more so with the advent of 
the COVID-19 pandemic. Several research studies have shown how the effects of 
fake news dissemination can be mitigated by promoting greater competence 
through 
lifelong learning and discussion communities, and generally rigorous training 
in the scientific method and broad interdisciplinary education. 
The urgent need for models that can describe the growing infodemic of fake news 
has been highlighted by the current pandemic. The resulting slowdown in 
vaccination campaigns due to misinformation and generally the inability of 
individuals to discern the reliability of information is posing enormous risks 
to the governments of many countries. In this research using the tools of 
kinetic theory we describe the interaction between fake news spreading and 
competence of individuals through multi-population models in which fake news 
spreads analogously to an infectious disease with different impact depending on 
the level of competence of individuals. The level of competence, in particular, 
is subject to an evolutionary dynamic due to both social interactions between 
agents and external learning dynamics. The results show how the model is able 
to correctly describe the dynamics of diffusion of fake news and the important 
role of competence in their containment.
\end{abstract}

{\bfseries Keywords}: fake news, compartmental models, competence, learning 
dynamics, 
interacting agents, socio-economic kinetic models, mean-field models 

%\tableofcontents

\section{Introduction}
With the rise of the Internet, connections among people has become easier than 
ever; so has been for the availability of information and its accessibility. As 
such, the Internet is also the source of unprecedented collective phenomena, 
some of which, however, cast shadows on our contemporary 
society^^>\cite{Bak-Colemane2025764118}. 
Indeed, the dissemination of heavily biased, or worse, downright false 
information, once relatively moderate in size, and limited possibly to the 
class of hoaxes and scams, exploded in the last decade, creating the broader 
category of^^>\emph{fake news}. The urgent need for models that can describe 
the increasing spread of fake news has been highlighted by the current COVID-19 
pandemic. Governments in many countries have found themselves in enormous 
difficulty because of the slowdown in vaccination campaigns due to the spread 
of false information and the inability of individuals to discern the 
authenticity of such information^^>\cite{fake1,fake2}.

Let us first briefly recall some of the main challenges in modeling fake news.
First of all, one of the priorities is to introduce a definition of 
fake news with a consensus that is wide enough to make research works 
relatable. In this direction, one of the most accepted (though not universally 
so) traits for fake news to be labeled so is 
\emph{purpose}: fake news is 
intentionally false 
news^^>\cite{Zhang2020,Shu2017FakeND,Allcott,Gelfert201884}. The concept of 
purpose seems to be useful when 
differentiating between theories that are focused on the content rather than on 
the conveyor^^>\cite{Zhang2020}. In this sense, a piece of information that is 
accidentally false (e.g., by inaccuracy) is substantially different, both 
semantically and stylistically, from a maliciously fabricated one.

Next, the challenge is to detect fake news. The majority of recent lines of 
research in the direction of automatic detection moves toward the aid of 
\emph{big data} and artificial intelligence 
tools^^>\cite{Shin2018278,Conroy20151,Ruchansky2017797,Vargo20182028}. An 
alternative strategy is instead \emph{component-based} and 
focuses on the analysis of the multiple parts involved in the diffusion of the 
fake news, that is, both on the side of the creator and on the side of the 
user, but also on the linguistics and semantics of the actual content and on 
its style, and finally on the social context of the information 
(see^^>\cite{Zhang2020} and references therein).

Different lines are possible, though. Information theory, for instance, has 
been used to model fake news: in^^>\cite{fninfoth}, fake news are defined as 
time series with an inherent bias, that is, its expectation is nonzero, 
involved in a stochastic process of which the user tries to judge the 
likelihood of the truth, together with noise.

Finally, epidemiological theory has been proving for long to be fertile ground 
for modeling of fake news^^>\cite{Daley19641118,daley_gani_1999}, 
especially in the somewhat 
broader category of rumor-spreading dynamics. 
The analogy between rumors and epidemics has often proved 
fruitful: Daley and Kendall^^>\cite{Daley19641118} took inspiration by the 
classical works of Kermack and Mckendrick^^>\cite{KermackACT,Hethcote2000TheMO} 
to propose a 
SIR-like model involving ignorant, spreader and stifler agents who played the 
role of the susceptible, infectious and recovered ones in^^>\cite{KermackACT}. 
Since the seminal paper^^>\cite{Daley19641118}, rumor-spreading dynamics has 
taken ideas from epidemiological models to improve their prediction accuracy.

Recently, networks theory 
delved in this direction, too^^>\cite{Cheng,networks,zhao2019fake,trammell}. 
Substantial research has merged networks and 
epidemiology through for instance classical compartmental 
models like the SIR (both epidemiological 
in^^>\cite{PhysRevLett.86.3200} and 
rumor-oriented^^>\cite{PIQUEIRA2020123406}), 
the SIS^^>\cite{PhysRevLett.89.108701} and the 
SIRS^^>\cite{PhysRevLett.86.2909,IEEE}. Moreover, more sophisticated 
epidemiological models have been developed, like the SEIZ 
model^^>\cite{BETTENCOURT2006513} describing  the evolution in time of the 
compartments of susceptible, exposed, infectious and skeptic agents, which has 
been adapted to the analysis of 
fake news dissemination (see, 
e.g.,^^>\cite{maleki2021using,rumorsontwitter}). In 
particular, the tendency seems to be to define the skeptic 
agents like the ones who are aware of the information but do not actively 
spread it^^>\cite{rumorsontwitter}. In a symmetric fashion, spreaders need not 
to believe a piece of 
information to be able to spread it (this is especially useful when thinking 
that bots are often encountered in social networks^^>\cite{datarepository}, 
both for legitimate and malicious purposes). This descriptions are also 
sensible in terms of matching the model with data available.

In this paper we follow this pathway: borrowing ideas from kinetic 
theory^^>\cite{PhysRevE.102.022303,intermultiagent}, we combine a classical 
compartmental approach inspired by 
epidemiology^^>\cite{Hethcote2000TheMO,KermackACT} with a kinetic description 
of the effects of competence^^>\cite{PARESCHI2017201,wealthPareschiToscani}.
We refer also to the recent work^^>\cite{rey} concerning evolutionary models
for knowledge.
In fact, the first wave of initiatives addressing fake news focused on news 
production by trying to limit citizen exposure to fake news. This can be done 
by fact-checking, labeling stories as fake, and eliminating them before they 
spread. Unfortunately, this strategy has already been proven not to work, it is 
indeed unrealistic to expect that only high quality, reliable information will 
survive. As a result, governments, international organizations, and social 
media companies have turned their attention to digital news consumers, and 
particularly children and young adults. From national campaigns in several 
countries to the OECD, there is a wave of action to develop new curricula, 
online learning tools, and resources that foster the ability to \lq\lq spot 
fake news\rq\rq^^>\cite{PIAAC19}. 

It is therefore of paramount importance to build models capable of describing 
the interplay between the dissemination of fake news and the creation of 
competence among the population. To this end, the approach we have followed in 
this paper falls within the recent socio-economic modeling described by kinetic 
equations (see^^>\cite{intermultiagent} for a recent monograph on the subject). 
More precisely, we adapted the competence model introduced 
in^^>\cite{PARESCHI2017201,wealthPareschiToscani} to a compartmental model 
describing fake news dissemination. Such a model allows not only to introduce 
competence as a static feature of the dynamics but as an evolutionary component 
both taking into account learning by interactions between agents and possible 
interventions aimed at educating individuals in the ability to identify fake 
news. Furthermore, in our modeling approach agents may have memory of fake news 
and as such be permanently immune to it once it has been detected, or fake news 
may not have any inherent peculiarities that would make it memorable enough for 
the population to immunize themselves against it in the future. The approach 
can be easily adapted to other compartmental models present in the literature, 
like the ones previously discussed^^>\cite{BETTENCOURT2006513, maleki2021using, 
PIQUEIRA2020123406}. 

The rest of the manuscript is organized as follows. In Section^^>\ref{sec:2} we 
introduce the structured kinetic model describing the spread of fake news in 
presence of different competence levels among individuals. The main properties 
of the resulting kinetic models are also analyzed. Next, Section^^>\ref{sec:3} 
is devoted to study the Fokker-Planck approximation of the kinetic model and to 
derive the corresponding stationary states in terms of competence. Several 
numerical results are then presented in Section^^>\ref{sec:4} that illustrate 
the theoretical findings and the capability of the model to describe transition 
effects in the spread of fake news due to the interaction between 
epidemiological and competence parameters. Some concluding remarks are reported 
in the last Section together with details on the 
theoretical results and the numerical methods in two separate appendices.

\section{Fake news spreading in a socially structured population}\label{sec:2}

In this section, we introduce a structured model for the dissemination of fake 
news in presence of different levels of skills among individuals in detecting 
the actual veracity of information, by combining a compartmental model in 
epidemiology and rumor-spreading 
analysis^^>\cite{Hethcote2000TheMO, daley_gani_1999} with the kinetic 
model of competence evolution proposed in^^>\cite{PARESCHI2017201}.

We consider a population of individuals divided into four classes.  The 
oblivious 
ones, still not aware of the news; the reflecting ones, who are aware of the 
news and 
are evaluating how to act; the spreader ones, who  
actively disseminate the news and the silent ones, who have recognized the fake 
news and do not contribute to its spread.
Terminology, when describing this compartmental models, is not fully 
established; however, the dominant one, inspired by epidemiology, refers to the 
definitions provided by 
Daley^^>\cite{daley_gani_1999} of a population composed of ignorant, spreader 
and stifler individuals. The class of reflecting agents can be referred to as 
a group that has a time-delay before taking a decision and enter an active 
compartment^^>\cite{BETTENCOURT2006513, maleki2021using}.

Notation, i.e., the choice of letters to represent the compartments, is even 
more scattered and somewhat confusing.  In 
Table^^>\ref{tab:notations} for readers' convenience we have summarized some of 
the different possible  
choices of letters and terminology found in literature.
Given the widespread use of epidemiological models compared to fake news 
models, in order to make the analogies easier to understand, we chose to align 
with notations conventionally used in epidemiology. Therefore, in 
the rest of the paper we will describe the population in terms of susceptible 
agents (S), who are the oblivious ones; exposed agents (E), who are in the 
time-delay compartment after exposure and before shifting into an active class; 
infectious agents (I), who are the spreader ones and finally removed agents (R) 
who are aware of the news but not actively engaging in its spread. 

Note that this subdivision of the population does not take into account actual 
beliefs of agents about the truth of the news, so that removed agents, for 
instance, need not be actually skeptic, nor the spreaders need to actually 
believe the news.  To simplify the mathematical treatment, as in the original 
works by Daley and Kendall^^>\cite{Daley19641118,daley_gani_1999}, we ignored 
the possible \lq active\rq\ effects of the population of removed individuals by 
interacting with other compartments and producing immunization among 
susceptible (the role of skeptic individuals 
in^^>\cite{BETTENCOURT2006513,maleki2021using}) and remission among spreaders 
(the role of stiflers in^^>\cite{PIQUEIRA2020123406}). Of course, the model 
easily generalizes to include these additional dynamics.

The main novelty in our approach is to consider an additional structure on the 
population based on the concept 
of competence of the agents, here understood as the ability to assess and 
evaluate information. 

\begin{table}
\begin{center}
\begin{tabular}{lllll}
\toprule
 SEIR (this paper) 
& DK^^>\cite{Daley19641118,daley_gani_1999} & ISR^^>\cite{PIQUEIRA2020123406} & 
SEIZR^^>\cite{BETTENCOURT2006513}
& SEIZ^^>\cite{maleki2021using}\\
\toprule
\multicolumn{5}{c}{\bf Category name}\\
\midrule
 Susceptible 
& Ignorant & Ignorant & Susceptible
& Susceptible\\
 Exposed  
& - & - & Idea incubator
& Exposed\\
 Infectious  
& Spreader & Spreader & Idea adopter
& Infectious\\
 Removed 
& Stifler & Stifler & Skeptic/Recovered
& Skeptic\\
\midrule
\multicolumn{5}{c}{\bf Variable notation}\\
\midrule
S  & X & I & S & S\\
E  & - & - & E & E\\
I  & Y & S & I & I\\
R  & Z & R & Z/R & Z\\
\bottomrule
\end{tabular}
\end{center}
\caption{Different compartments and notations for some of the models found in 
literature.}
\label{tab:notations}
\end{table}

Let us suppose that agents in the system are completely characterized by their 
competence \mbox{$x \in X \contenutoin \R^+$}, measured in a suitable unit. We 
denote 
by \mbox{$f_S = f_S(x, t)$}, \mbox{$f_E = f_E(x,t)$}, \mbox{$f_I = f_I(x,t)$}, 
\mbox{$f_R = f_R(x,t)$}, the competence 
distribution at time^^>\mbox{$t > 0$} of susceptible, exposed, infectious and 
removed individuals, respectively. Aside from natality or 
mortality concerns (i.e., the social network is a closed system---nobody enters 
or leaves it during the diffusion of the fake news, which is a common 
assumption, based on the average lifespan of fake news) we therefore have:
\[
\int_X \bigl(f_S(x,t) + f_E(x,t) + f_I(x,t) + f_R(x,t) \bigr)\, \de x = 1,\quad 
t > 0,
\]
which implies that we will refer to
\begin{align*}
S(t) &= \int_X f_S(x,t)\, \de x, &
E(t) &= \int_X f_E(x,t)\, \de x,\\
I(t) &= \int_X f_I(x,t)\, \de x, &
R(t) &= \int_X f_R(x,t)\, \de x
\end{align*}
as the fractions of the population that are susceptible, exposed, infected, or 
recovered 
respectively. We also denote the relative mean competences as
\begin{align*}
m_S(t) &= \int_X x f_S(x,t)\, \de x, &
m_E(t) &= \int_X x f_E(x,t)\, \de x, \\
m_I(t) &= \int_X x f_I(x,t)\, \de x, &
m_R(t) &= \int_X x f_R(x,t)\, \de x.
\end{align*}

\subsection{A SEIR model describing fake news dynamics}
The fake news dynamics proceeds as follows: a susceptible agent gets to know it 
by a spreader. At this point, the now-exposed agent evaluates the piece of 
information---the reflecting, or delay, stage---and decide whether to
share it with other individuals (and turning into a spreader themselves) or to 
keep silent, removing themselves by the dissemination process.

\begin{figure}
\centering
\includegraphics[scale=1.5]{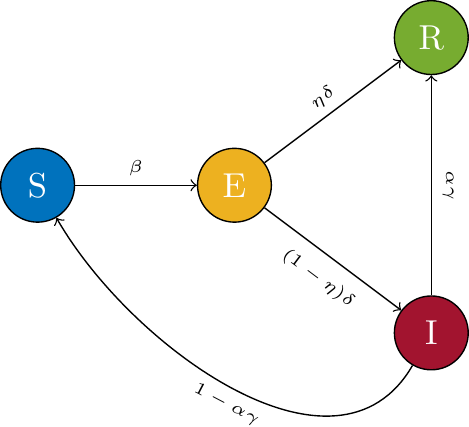}
\caption{SEIR diagram with transition rates.}
\label{fig:SEIRdiagram}
\end{figure}
When the dynamic is independent from the knowledge of individuals, the model 
can be expressed by 
the following system of ODEs
\begin{equation}
\left\lbrace
\begin{aligned}
\der St &= -\beta SI + (1 - \alpha)\gamma I\\ 
\der Et &= \beta SI - \delta E \\
\der It &= (1-\eta)\delta E - \gamma I,\\
\der Rt &= \eta\delta E + \alpha\gamma I,
\end{aligned}
\right.
\label{eq:seirnormale}
\end{equation}
with $S+E+I+R=1$ and where $\beta$ is the contact rate between the class of the 
susceptible and the 
class of infectious, $\delta$ is the rate at which agents make their decision 
about spreading the news or not, $1 - \eta$ is the portion of agents who become 
infectious and $\gamma$ is the rate at which spreaders remove themselves from 
the compartment, due, e.g., to loss of interest in sharing the news or 
forgetfulness. Finally, $\alpha$ is related to the specificity of the fake news 
and the probability of individulas to remember it.  A probability of^^>$0$ 
means that the fake news has not 
any inherent peculiarity (e.g., in terms of content, structure, style, \ldots) 
that can make it memorable enough for the population to \lq immunize\rq\ 
against it in the future, while a probability of^^>$1$ allows for the agents to 
have the full ability to not fall for that fake news a second time. The various 
parameters have been summarized in Table \ref{tab:parameters}. The diagram of 
the SEIR model^^>\eqref{eq:seirnormale} is shown in 
Figure^^>\ref{fig:SEIRdiagram}.
\begin{table}
\begin{center}
\begin{tabular}{cl}
\toprule
 Parameter & Definition\\
\toprule
$\beta$ & contact rate between susceptible and infected individuals\\
$1/\delta$ & average decision time on whether or not to spread fake news\\
$\eta$ & probability of deciding not to spread fake news \\
$1/\gamma$ & average duration of a fake news\\
$\alpha$ & probability of remembering fake news\\
\bottomrule
\end{tabular}
\end{center}
\caption{Parameters definition in the SEIR model \eqref{eq:seirnormale}.}
\label{tab:parameters}
\end{table}
It is straightforward to notice that when^^>$\alpha$ and $\eta$ are zero, 
system^^>\eqref{eq:seirnormale} specializes in a classic SEIS epidemiological 
model. This is consistent with treating the dissemination of non-specific fake 
news in a population as the spread of a disease with multiple strains, for 
which a durable immunization is never attained. In this case 
system^^>\eqref{eq:seirnormale} has two equilibrium states: a 
disease-free equilibrium state^^>$(1,0,0)$ and an endemic equilibrium 
state^^>\mbox{$\tilde P = (\tilde S, \tilde E, \tilde I)$} where
\begin{equation}\label{eq:endemicequilibria}
\tilde S = \frac1{R_0}, \qquad
\tilde E = \frac{\gamma}{\gamma + \delta}\biggl(1 - \frac1{R_0}\biggr),\qquad
\tilde I = \frac{\delta}{\gamma + \delta}\biggl(1 - \frac1{R_0}\biggr),
\end{equation}
and $R_0 = \beta/\gamma$ is the basic reproduction number. It is 
known^^>\cite{korobeinikov} that if 
\mbox{$R_0 > 1$} the endemic equilibrium state^^>$\tilde P$ of 
system^^>\eqref{eq:seirnormale} is globally asymptotically stable.

If instead $\alpha > 0$ or $\eta > 0$, there also is the possibility to 
permanently immunize against fake news with those traits; moreover, both 
infectious and exposed agents eventually vanish, leaving only the 
susceptible and removed compartments populated. In the case of maximum 
specificity of the fake news, i.e., \mbox{$\alpha = 1$}, the stationary 
equilibrium state has the form
\begin{equation}\label{eq:orsmequilibria}
S(t) \to S^\infty, \quad E(t) \to 0, \quad I(t) \to 0, \quad R(t) \to R^\infty 
= 1-S^\infty,
\end{equation}
where $S^\infty$ is solution of the nonlinear equation
\begin{equation}
\log\frac{S_0}{S^\infty} = \frac\beta\gamma(1 - \eta)(1 - S^\infty),
\end{equation}
in which $S_0$ is the initial datum $S(t = 0)$. 

We refer to^^>\cite{BETTENCOURT2006513,maleki2021using,PIQUEIRA2020123406} for 
the inclusion of additional interaction dynamics, taking into account 
counter-information effects due to the removed population interacting against 
susceptible and infectious, and the relative analysis of the resulting 
equilibrium states. 

\subsection{The interplay with competence and learning}
In the following, we combine the evolution of the densities according to the 
SEIR model \eqref{eq:seirnormale} with the competence dynamics proposed 
in^^>\cite{PARESCHI2017201}. 
We refer to the degree of competence 
that an individual can gain or loose in a single interaction from 
the background as \mbox{$z \in \R^+$}; in what follows we denote by^^>$C(z)$ 
the bounded-mean distribution of $z$, satisfying
\[
\int_{\R^+} C(z)\, \de z = 1, \quad \int_{\R^+} zC(z)\, \de z = m_B.
\]
Assuming a susceptible agent has a competence level $x$ and interacts with 
another one belonging to the various compartments in the population and having 
a competence level^^>$x\*$, their levels after the interaction will be given by
\begin{equation}
\begin{aligned}
\left\lbrace
\begin{aligned}
x'   &= (1 - \lambda_S(x))x + \lambda_{CJ}(x)x\* + \lambda_{BS}(x)z
        + \kappa_{SJ} x\\
x\*' &= (1 - \lambda_J(x\*))x\* + \lambda_{CS}(x\*)x + \lambda_{BJ}(x\*)z
        + \tilde\kappa_{SJ} x\*,
\end{aligned}
\right.
     && J \in \ins{S, E, I, R}
\end{aligned}
\label{eq:competencebinaryS}
\end{equation}
where $\lambda_S(\cdot)$ and $\lambda_{BS}(\cdot)$ quantify the 
amount of competence lost by susceptible individuals by the natural process of 
forgetfulness and the amount gained by susceptible individuals from the 
background, respectively. $\lambda_{CJ}$, instead, models the competence 
gained through the interaction with members of the 
class^^>$J$, with \mbox{$J \in \ins{S, E, I, R}$}; a possible choice 
for^^>$\lambda_{CJ}(x)$ is^^>\mbox{$\lambda_{CJ}(x) = \lambda_{CJ}\chi(x \ge 
\bar x)$}, where $\chi(\cdot)$ is the characteristic function and \mbox{$\bar x 
\in X$} a minimum level of competence required to the agents for increasing 
their own skills by interactions. Finally, $\kappa_{SJ}$ and 
$\tilde\kappa_{SJ}$ are independent and identically distributed zero-mean 
random variables with the same variance^^>$\sigma(t)$ to consider the 
non-deterministic nature of the competence acquisition process. 

The binary interactions involving the exposed agents can be similarly defined
\begin{equation}
\begin{aligned}
\left\lbrace
\begin{aligned}
x'   &= (1 - \lambda_E(x))x + \lambda_{CJ}(x)x\* + \lambda_{BE}(x)z
        + \kappa_{EJ} x\\
x\*' &= (1 - \lambda_J(x\*))x\* + \lambda_{CE}(x\*)x + \lambda_{BJ}(x\*)z
        + \tilde\kappa_{EJ} x\*,
\end{aligned}
\right.
     && J \in \ins{S, E, I, R}
\end{aligned}
\label{eq:competencebinaryE}
\end{equation}
the same holds for the interactions concerning the infectious fraction of 
the population
\begin{equation}
\begin{aligned}
\left\lbrace
\begin{aligned}
x'   &= (1 - \lambda_I(x))x + \lambda_{CJ}(x)x\* + \lambda_{BI}(x)z
        + \kappa_{IJ} x\\
x\*' &= (1 - \lambda_J(x\*))x\* + \lambda_{CI}(x\*)x + \lambda_{BJ}(x\*)z
        + \tilde\kappa_{IJ} x\*,
\end{aligned}
\right.
     && J \in \ins{S, E, I, R}
\end{aligned}
\label{eq:competencebinaryI}
\end{equation}
and finally we have the interactions regarding the removed agents 
\begin{equation}
\begin{aligned}
\left\lbrace
\begin{aligned}
x'   &= (1 - \lambda_R(x))x + \lambda_{CJ}(x)x\* + \lambda_{BR}(x)z
        + \kappa_{RJ} x\\
x\*' &= (1 - \lambda_J(x\*))x\* + \lambda_{CR}(x\*)x + \lambda_{BJ}(x\*)z
        + \tilde\kappa_{RJ} x\*,
\end{aligned}
\right.
     && J \in \ins{S, E, I, R}.
\end{aligned}
\label{eq:competencebinaryR}
\end{equation}

It is reasonable to assume that both the processes of gain and loss of 
competence from the interaction with other agents or with the background 
in^^>\eqref{eq:competencebinaryS}--\eqref{eq:competencebinaryR} are bounded by 
zero. Therefore we suppose that if \mbox{$J, H \in \ins{S, E, I, R}$}, and if 
\mbox{$\lambda_J \in [\lambda_J^-, \lambda_J^+]$}, with \mbox{$\lambda_J^- > 0$}
and \mbox{$\lambda_J^+ < 1$}, and \mbox{$\lambda_{CJ}(x),\lambda_{BJ}(x) 
\in [0, 1]$} then $\kappa_{HJ}$ may, for example, be uniformly distributed 
in^^>\mbox{$[-1 + \lambda_J^+, 1 - \lambda_J^+]$}.

In order to combine the compartmental model SEIR with the evolution 
of the competence levels given by 
equations^^>^^>\eqref{eq:competencebinaryS}--\eqref{eq:competencebinaryR} we 
introduce the interaction operator^^>$Q_{HJ}(\cdot, \cdot)$ following the 
standard Boltzmann-type theory^^>\cite{intermultiagent}. As earlier, we will 
denote with^^>$J$ a suitable compartment of 
the population, i.e., \mbox{$H, J \in \ins{S, E, I, R}$}, and we will use the 
brackets^^>$\expvalue\cdot$ to indicate the expectation with respect to the 
random variable^^>$\kappa_{HJ}$. Thus, if $\psi(x)$ is an observable function, 
then the action of^^>$Q_{HJ}(f_H,f_J)(x,t)$ on^^>$\psi(x)$ is given by
\begin{equation}\label{eq:wk1}
\int_{\R^+} Q_{SJ}(f_S,f_J) \psi(x)\, \de x = \expvalue*{
                \int_{\R^2_+} f_S(x,t) f_J(x\*,t)
                \bigl(\psi(x') - \psi(x)\bigr)\, \de x\*\de x
},
\end{equation}
with $x'$ defined by^^>\eqref{eq:competencebinaryS}
\begin{equation}\label{eq:wk2}
\int_{\R^+} Q_{EJ}(f_E,f_J) \psi(x)\, \de x = \expvalue*{
                \int_{\R^2_+} f_E(x,t) f_J(x\*,t)
                \bigl(\psi(x') - \psi(x)\bigr)\, \de x\*\de x
},
\end{equation}
with $x'$ defined by^^>\eqref{eq:competencebinaryE}
\begin{equation}\label{eq:wk3}
\int_{\R^+} Q_{IJ}(f_I,f_J) \psi(x)\, \de x = \expvalue*{
                \int_{\R^2_+} f_I(x,t) f_J(x\*,t)
                \bigl(\psi(x') - \psi(x)\bigr)\, \de x\*\de x
},
\end{equation}
with $x'$ defined by^^>\eqref{eq:competencebinaryI},
\begin{equation}\label{eq:wk4}
\int_{\R^+} Q_{RJ}(f_R,f_J) \psi(x)\, \de x = \expvalue*{
                \int_{\R^2_+} f_R(x,t) f_J(x\*,t)
                \bigl(\psi(x') - \psi(x)\bigr)\, \de x\*\de x
},
\end{equation}
with $x'$ defined by^^>\eqref{eq:competencebinaryR}. All the above operators 
preserve the total number of agents as the unique interaction invariant, 
corresponding to $\psi(\cdot) \equiv 1$. 

The system then reads:
\begin{equation}
\left\lbrace
\begin{aligned}
\pd{f_S(x,t)}{t}  &= -K(x,t)f_S(x,t) + (1-\alpha(x))\gamma(x)f_I(x,t)
                     + \sum_{\mathclap{J \in \ins{S,E,I,R}}} 
                     Q_{SJ}(f_S,f_J)(x,t),\\ 
\pd{f_E(x,t)}{t}  &= K(x,t)f_S(x,t) - \delta(x) f_E(x,t)
                     + \sum_{\mathclap{J \in \ins{S,E,I,R}}}
                     Q_{EJ}(f_E, f_J)(x,t),\\
\pd{f_I(x,t)}{t}  &= \delta(x)(1 - \eta(x)) f_E(x,t) - \gamma(x)f_I(x,t)
                     + \sum_{\mathclap{J \in \ins{S,E,I,R}}}
                     Q_{IJ}(f_I, f_J)(x,t),\\
\pd{f_R(x,t)}{t}  &= \delta(x)\eta(x) f_E(x,t) + \alpha(x)\gamma(x) f_I(x,t)
                     + \sum_{\mathclap{J \in \ins{S,E,I,R}}}
                     Q_{RJ}(f_R, f_J)(x,t),
\end{aligned}
\right.
\label{eq:seiscompetenza1}
\end{equation}
where the function
\[
K(x,t) = \int_{\R^+} \beta(x,x\*) f_I(x\*,t)\, \de x\*
\]
is responsible for the contagion, $\beta(x,x\*)$ being the contact rate between 
agents with competence levels^^>$x$ and^^>$x\*$. In the above formulation we 
also assumed $\beta$, 
$\gamma$, $\delta$, $\eta$ and^^>$\alpha$ functions of^^>$x$. Note that, 
clearly, the most important parameters influenced by individuals' competence 
are $\beta(x,x\*)$, since individuals have the highest rates of contact with 
people belonging to the same social class, and thus with a similar level of 
competence, $\delta(x)$ as individuals with greater competence invest more time
in checking the authenticity of information,  
and $\eta(x)$, which characterizes individuals' decision to spread fake news. 
On the other hand, the values of $\gamma$  and $\alpha$ 
we may assume to be less influenced by the level of expertise of individuals.  

\subsection{Properties of the kinetic SEIR model with competence}
In this section we analyze some of the properties of the Boltzmann 
system^^>\eqref{eq:seiscompetenza1}. First let us consider the reproducing 
ratio in presence of knowledge.

By integrating system \eqref{eq:seiscompetenza1} against $x$, and considering 
only the compartments of individuals which may disseminate the fake news we have
\begin{equation}
\left\lbrace
\begin{aligned}
\der {E(t)}t  &= \int_X K(x,t)f_S(x,t)\,dx - \int_X \delta(x) f_E(x,t)\,dx,\\
\der {I(t)}t  &= \int_X \delta(x)(1 - \eta(x)) f_E(x,t)\,dx - \int_X 
\gamma(x)f_I(x,t)\,dx.\\
\end{aligned}
\right.
\label{eq:seisintegrated}
\end{equation} 
In the above derivation we used the fact that the Boltzmann interaction terms 
describing knowledge evolution among agents preserve the total number of 
individuals and therefore vanish. Following the analysis in 
\cite{BertagliaPareschi}, and omitting the details for brevity, we obtain a 
reproduction number generalizing the classical one
\begin{equation}\label{eq:reproductionnumber}
R_0(t) = \frac{\int_X K(x,t)f_S(x,t)\,dx}{\int_X \gamma(x)f_I(x,t)\,dx}.
\end{equation}

Next, following \cite{PhysRevE.102.022303}, we can prove uniqueness of the 
solution of^^>\eqref{eq:seiscompetenza1} in the simplified case of constant 
parameters: \mbox{$\beta(x,x\*) = \beta > 0$}, \mbox{$\gamma(x) = \gamma > 0$}, 
\mbox{$\delta(x) = \delta > 0$}, \mbox{$\eta(x) = \eta \in [0,1]$}, 
\mbox{$\alpha(x) = \alpha \in [0,1]$}. In this 
case, exploiting the fact that the interaction operator^^>$Q(\cdot,\cdot)$ has a 
natural connection with the Fourier transform by choosing its 
kernel^^>\mbox{$\e^{-ix\xi}$} as test function, we can analyze the 
system^^>\eqref{eq:seiscompetenza1} with the Fourier transforms of the 
densities as unknowns. 

Indeed, given a function $f(x) \in L_1(\R^+)$, its Fourier transform is defined 
as
\[
\ft f(\xi) = \int_\R \e^{-ix\xi} f(x)\, \de x.
\] 
The system^^>\eqref{eq:seiscompetenza1} becomes
\begin{equation}
\left\lbrace
\begin{aligned}
\pd{\ft f_S(\xi,t)}{t}  &= -\beta I(t) \ft f_S(\xi,t) + (1-\alpha) \gamma
                           \ft f_I(\xi,t)
                           + \sum_{\mathclap{J \in \ins{S,E,I,R}}}
                             \ft Q_{SJ}(\ft f_S, \ft f_J)(\xi,t),\\ 
\pd{\ft f_E(\xi,t)}{t}  &= \beta I(t) \ft f_S(\xi,t) - \delta \ft f_E(\xi,t)
                           + \sum_{\mathclap{J \in \ins{S,E,I,R}}}
                             \ft Q_{EJ}(\ft f_E, \ft f_J)(\xi,t),\\
\pd{\ft f_I(\xi,t)}{t}  &= \delta(1 - \eta) \ft f_E(\xi,t)
                           - \gamma \ft f_I(\xi,t)
                           + \sum_{\mathclap{J \in \ins{S,E,I,R}}}
                             \ft Q_{IJ}(\ft f_I, \ft f_J)(\xi,t),\\
\pd{\ft f_R(\xi,t)}{t}  &= \delta\eta \ft f_E(\xi,t)
                           + \alpha\gamma \ft f_I(\xi,t)
                           + \sum_{\mathclap{J \in \ins{S,E,I,R}}}
                             \ft Q_{RJ}(\ft f_R, \ft f_J)(\xi,t),
\end{aligned}
\right.
\label{eq:seiscompetenzaFourier}
\end{equation}
where the operators $\ft Q_{HJ}(\ft f_H, \ft f_J)$ are defined in terms of the 
Fourier transforms of their arguments for^^>\mbox{$J \in \ins{S, E, I, R}$}, so 
that
\[
\ft Q_{HJ}(\ft f_H, \ft f_J) = \expvalue{\ft f_H(A_{HJ} \xi -\lambda_{BH}z, t)}
                          \ft f_J(\lambda_{CJ}\xi,t) - \ft f_H(\xi,t) J(t),
\]
where $A_{HJ}$, with $H,J\in \ins{S, E, I, R}$ is defined as
\begin{equation}\label{eq:defAHJ}
A_{HJ} = 1 - \lambda_H +\kappa_H.
\end{equation}
We suppose that the parameters satisfy the condition
\begin{equation}\label{eq:nucondition}
\nu = \max_{H,J \in \ins{S, E, I, R}} [\lambda_{CJ}^2 + \expvalue{A_{JH}^2}] < 
1,
\end{equation}
which will prove useful in the proof. 

As in^^>\cite{PhysRevE.102.022303} we recall a class of metrics which is of 
natural use in bilinear Boltzmann equations^^>\cite{intermultiagent}. Let $f$ 
and^^>$g$ be probability densities. Then, for^^>\mbox{$s > 0$} we define
\begin{equation}\label{eq:defd2}
d_s(f,g) = \sup_{\xi \in \R}\frac{\abs{\ft f(\xi) - \ft g(\xi)}}{\abs \xi ^s},
\end{equation}
which is finite whenever $f$ and $g$ have equal moments up to the integer part 
of^^>$s$ or to^^>\mbox{$s-1$} if $s$ is an integer.

We have the following result.
\begin{theorem}\label{teo:fouriermetric}
Let $J \in \ins{S, E, I, R}$, and let $f_J(x,t)$ and^^>$g_J(x,t)$ be two 
solutions 
of the system^^>\eqref{eq:seiscompetenza1} with initial values $f_J(x,0)$ 
and^^>$g_J(x,0)$ such that $d_2(f_J(x,0), g_J(x,0)$ is finite. Then, 
condition^^>\eqref{eq:nucondition} implies that the Fourier based 
distance^^>$d_2(f_J(x,t), g_J(x,t)$ decays exponentially (in time) to zero, so 
that
\[
\sum_{\mathclap{J \in \ins{S,E,I,R}}} d_2(f_J(x,t), g_J(x,t) \le
\sum_{\mathclap{J \in \ins{S,E,I,R}}} d_2(f_J(x,0), g_J(x,0)\e^{-(1 - \nu)t}.
\]
\end{theorem}
For the details of the proof we refer to Appendix^^>\ref{appendix:A}.

\section{Mean-field approximation}\label{sec:3}
A highly useful tool to obtain information analytically on the large-time 
behavior of Boltzmann-type models are scaling techniques; in particular the 
so-called \textit{quasi-invariant} limit^^>\cite{intermultiagent}, which allows to derive the 
corresponding mean-field description of the kinetic 
model \eqref{eq:seiscompetenza1}.

Indeed, let us consider the case in which the interactions between agents 
 produce small variations of the competence. We scale the quantities involved 
in the binary interactions \eqref{eq:competencebinaryS}-\eqref{eq:competencebinaryR} accordingly
\begin{equation}\label{eq:quasiinvariant1}
\begin{aligned}
\lambda_{CJ} \to \epsilon\lambda_{CJ}, &&
\lambda_{BJ} \to \epsilon\lambda_{BJ}, &&
\lambda_J \to \epsilon\lambda_J,       &&
\sigma \to \epsilon\sigma,
\end{aligned}
\end{equation}
where $J\in \ins{S, E, I, R}$ and the functions involved in the dissemination 
of the fake news, as well
\begin{equation}\label{eq:quasiinvariant2}
\begin{aligned}
\beta(x,x\*) \to \epsilon\beta(x,x\*), &&
\gamma(x) \to \epsilon\gamma(x),       &&
\delta(x) \to \epsilon\delta(x),       &&
\eta(x) \to \epsilon\eta(x).
\end{aligned}
\end{equation}
We denote by $Q^\epsilon_{HJ}(\cdot, \cdot)$ the scaled interaction terms. 
Omitting 
the dependence on time on mean values and re-scaling time as 
\mbox{$t \to t/\epsilon$}, we obtain up to $\mathcal O(\epsilon)$
\[
\begin{aligned}
\frac1\epsilon \int_{\R^+} Q_{SJ}^{\epsilon}(f_S, f_J) \psi(x)\, \de x
              &\approx \int_{\R^+}
                 \biggl[
                 -\psi(x)'(\lambda_S xJ - \lambda_{CJ} m_J - \lambda_{BS} m_B J)
                 + \frac\sigma2 \psi(x)'' x^2 J
                 \biggr]
                 f_S(x,t)\, \de x\\
\frac1\epsilon \int_{\R^+} Q_{EJ}^{\epsilon}(f_E, f_J) \psi(x)\, \de x
              &\approx \int_{\R^+}
                 \biggl[
                 -\psi(x)'(\lambda_E xJ - \lambda_{CJ} m_J - \lambda_{BE} m_B J)
                 + \frac\sigma2 \psi(x)'' x^2 J
                 \biggr]
                 f_E(x,t)\, \de x\\
\frac1\epsilon \int_{\R^+} Q_{IJ}^{\epsilon}(f_I, f_J) \psi(x)\, \de x
              &\approx \int_{\R^+}
                 \biggl[
                 -\psi(x)'(\lambda_I xJ - \lambda_{CJ} m_J - \lambda_{BI} m_B J)
                 + \frac\sigma2 \psi(x)'' x^2 J
                 \biggr]
                 f_I(x,t)\, \de x\\
\frac1\epsilon \int_{\R^+} Q_{RJ}^{\epsilon}(f_R, f_J) \psi(x)\, \de x
              &\approx \int_{\R^+}
                 \biggl[
                 -\psi(x)'(\lambda_R xJ - \lambda_{CJ} m_J - \lambda_{BR} m_B J)
                 + \frac\sigma2 \psi(x)'' x^2 J
                 \biggr]
                 f_R(x,t)\, \de x,
\end{aligned}
\]
where we used a Taylor expansion for small values of $\varepsilon$ of 
\[
\psi(x')=\psi(x)+(x'-x)\psi'(x)+\frac{(x'-x)^2}{2}\psi''(x)+\mathcal 
O(\epsilon^2)
\]
in \eqref{eq:wk1}-\eqref{eq:wk4} and the scaled interaction rules 
\eqref{eq:competencebinaryS}-\eqref{eq:competencebinaryR}.

\subsection{Stationary solutions of Fokker-Planck SEIR models}
Let us impose that $\epsilon \to 0$, following^^>\cite{intermultiagent} from the computations of the previous section we formally obtain the Fokker-Planck 
system
\begin{align}
\pd{f_S(x,t)}{t} &= -K(x,t)f_S(x,t) + (1-\alpha(x))\gamma(x)f_I(x,t)
                    + \pd{}{x}[(x\lambda_S -\overline{m}(t)
                                           -\lambda_{BS}m_B)f_S(x,t)]\notag\\
                 &\phantom{=} {}+ \frac\sigma2 
                 \pd{^2}{x^2}(x^2f_S(x,t))\label{eq:fokkerplanck4}\\
\pd{f_E(x,t)}{t} &= K(x,t)f_S(x,t) -\delta(x)f_E(x,t)
                    + \pd{}{x}[(x\lambda_E -\overline{m}(t)
                                           -\lambda_{BE}m_B)f_E(x,t)]\notag\\
                 &\phantom{=} {}+ \frac\sigma2 
                 \pd{^2}{x^2}(x^2f_E(x,t))\label{eq:fokkerplanck5}\\
\pd{f_I(x,t)}{t} &= \delta(x)(1 - \eta(x))f_E(x,t) - \gamma(x)f_I(x,t)
                    + \pd{}{x}[(x\lambda_I -\overline{m}(t)
                                           -\lambda_{BI}m_B)f_I(x,t)]\notag\\
                 &\phantom{=} {}+ \frac\sigma2 
                 \pd{^2}{x^2}(x^2f_I(x,t))\label{eq:fokkerplanck6}\\
\pd{f_R(x,t)}{t} &= \delta(x)\eta(x)f_E(x,t) + \alpha(x)\gamma(x)f_I(x,t)
                    + \pd{}{x}[(x\lambda_R -\overline{m}(t)
                                           -\lambda_{BR}m_B)f_R(x,t)]\notag\\
                 &\phantom{=} {}+ \frac\sigma2 
                 \pd{^2}{x^2}(x^2f_R(x,t))\label{eq:fokkerplanck7}
\end{align}
where now
\[
\overline{m}(t) = \lambda_{CS}m_S(t) + \lambda_{CE}m_E(t) +
                    \lambda_{CI}m_I(t) + \lambda_{CR}m_R(t).
\]
We can consider the mean values system associated 
to^^>\eqref{eq:fokkerplanck4}--\eqref{eq:fokkerplanck7} in the case of constant 
epidemiological parameters
\begin{align}
\der{m_S(t)}{t} &= -\beta I(t)m_S(t) + (1-\alpha)\gamma m_I(t)
                   + \lambda S(t)(m(t)  - m_B)/2-\lambda 
                   m_S(t)\label{eq:fokkerplancmeank4}\\
\der{m_E(t)}{t} &= \beta I(t)m_S(t) -\delta m_E(t)
                   + \lambda E(t)(m(t)  - m_B)/2-\lambda 
                   m_E(t)\label{eq:fokkerplanckmean5}\\
\der{m_I(t)}{t} &= \delta(1 - \eta)m_E(t) - \gamma m_I(t)
                   + \lambda I(t)(m(t)  - m_B)/2-\lambda 
                   m_I(t)\label{eq:fokkerplanckmean6}\\
\der{m_R(t)}{t} &= \delta\eta m_E(t) + \alpha\gamma m_I(t)
                   + \lambda R(t)(m(t)  - m_B)/2-\lambda 
                   m_R(t)\label{eq:fokkerplanckmean7},
\end{align}
with
\[
m(t)=m_S(t) + m_E(t) + m_I(t) + m_R(t).
\]
In the case $\alpha > 0$ or $\eta > 0$, we know that $E(t) \to 0$, $I(t) \to 
0$, $S(t) \to S^\infty$ and $R(t) \to R^\infty = 1 - S^\infty$ due to mass 
conservation, so that $m_E(t) \to 0$ and $m_I(t) \to 0$ as well. Thus, adding 
all the equations together leads us to
\begin{align}\label{eq:meanconvergence2}
\frac\lambda2(m_S^\infty + m_R^\infty) + \frac\lambda2 m_B
&= \lambda(m_S^\infty + m_R^\infty),
\end{align}
i.e., $m_S^\infty + m_R^\infty = m_B$. At this point, adding together 
equations^^>\eqref{eq:fokkerplanck4} to^^>\eqref{eq:fokkerplanck7} gives us
\[
0 = \lambda\pd{}{x}(x - m_B)
            \sum_{\mathclap{J \in \ins{S,R}}}f_J^\infty(x) 
   + \frac\sigma2 \pd{^2}{x^2}\biggl(x^2
\sum_{\mathclap{\quad J \in \ins{S,R}}}f_J^\infty(x) \biggr),
\]
which has as solution an inverse Gamma density
\begin{equation}\label{eq:seirFPfinal}
f^\infty(x) = f_S^\infty(x) + f_R^\infty(x) =
              \frac{k^\mu}{\Gamma(\mu)}\frac{\e^{-k/x}}{x^{1 + \mu}},
\end{equation}
\[
\begin{aligned}
\mu = 1 + \frac{2\lambda}\sigma, &&\quad k = (\mu - 1)m_B.
\end{aligned}
\]
It is straightforward to see that the scaled Gamma densities
\[
f_S^\infty (x) = S^\infty \frac{k^\mu}{\Gamma(\mu)}
                                 \frac{\e^{-k/x}}{x^{1 + \mu}}
\quad
f_R^\infty (x) = (1-S^\infty) \frac{k^\mu}{\Gamma(\mu)}
                                 \frac{\e^{-k/x}}{x^{1 + \mu}}
\]
are solutions of the 
system^^>\eqref{eq:fokkerplanck4}--\eqref{eq:fokkerplanck7}.

If, instead, $\alpha = \eta = 0$, we find again the same solution 
as^^>\eqref{eq:seirFPfinal}, but in this case $J \to \tilde J$, where $\tilde 
J$ are defined as in^^>\eqref{eq:endemicequilibria}.

\begin{figure}
\centering
\includegraphics[width=0.5\columnwidth]{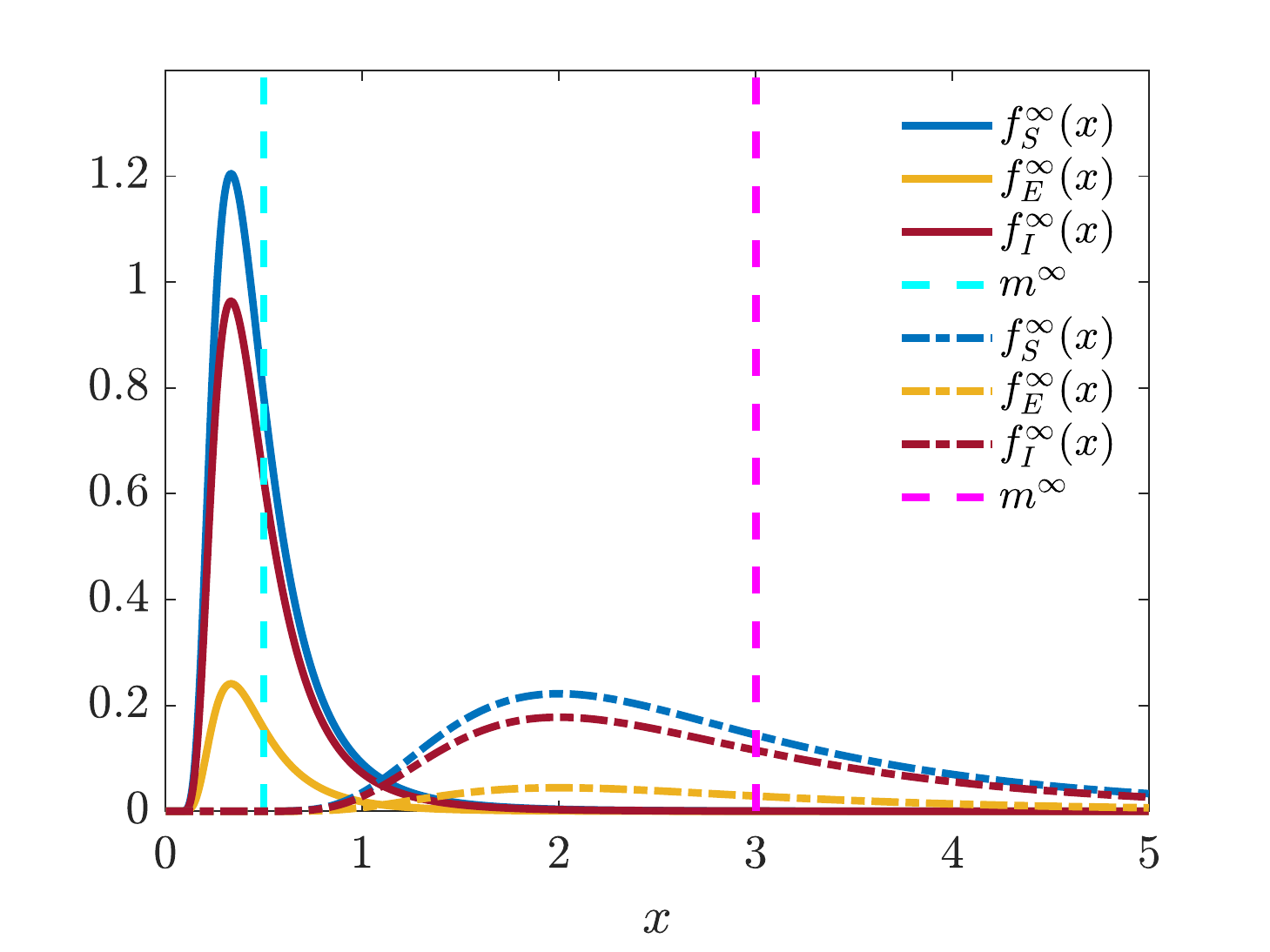}%
\hskip-.1cm
\includegraphics[width=0.5\columnwidth]{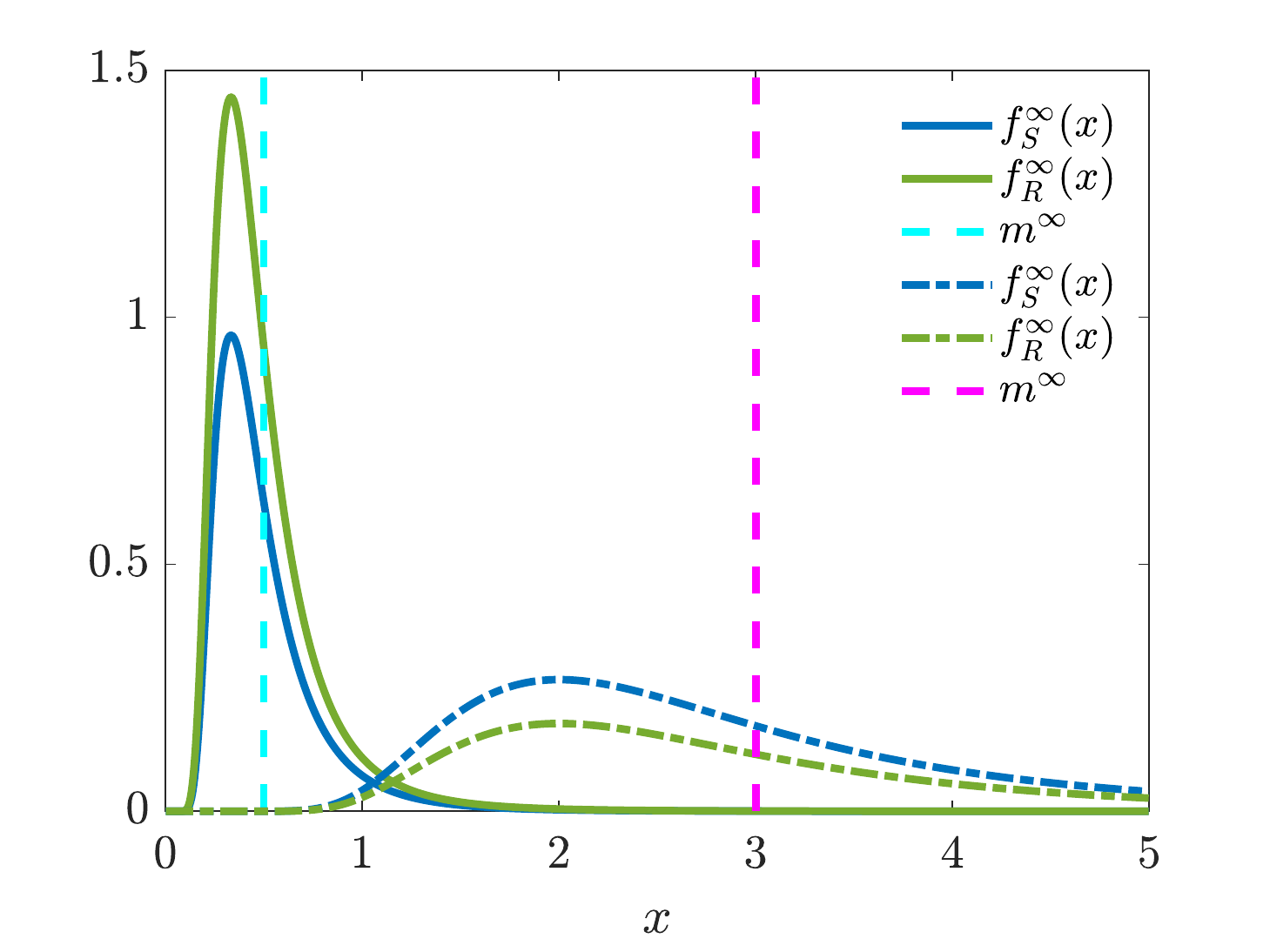}
\caption{Exact solutions for competence distributions at the end of 
epidemic\eqref{eq:seirFPfinal} for $\lambda = 0.1$, $\mu = 5$, $m_B = 0.5$ 
(solid), $m_B = 3$ (dash-dotted), $\tilde S = 0.5$, $\tilde E = 0.1$, $\tilde I 
= 0.4$ (left) and $S^\infty = 0.6$, $R^\infty= 0.4$ (right). 
Left: case $\alpha=\eta=0$; right: case $\alpha=0.2$, $\eta = 0.1$.}
\label{fig:stationarysols0125075}
\end{figure}

In Figure^^>\ref{fig:stationarysols0125075} we report two examples of 
the stationary solutions where we chose the competence variable^^>$z$ to be 
uniformly distributed in $[0,1]$: in the first case (left) we considered 
$\alpha = \eta = 0$, while in the second case (right) we set $\alpha = 0.2$ and 
$\eta = 0.1$.

\section{Numerical examples}\label{sec:4}

In this section we present some numerical tests to show the characteristics of 
the model in describing the dynamics of fake news dissemination in a population 
with a competence-based structure.

To begin with, we validate the Fokker-Planck model obtained as the 
\emph{quasi-invariant limit} of the Boltzmann equation: we will do so through a Monte 
Carlo method for the competence distribution^^>(see \cite{intermultiagent}, Chapter 5 for more details). 
Next, we approximate the
Fokker-Planck systems^^>\eqref{eq:fokkerplanck4}--\eqref{eq:fokkerplanck7} by 
generalizing the structure-preserving 
numerical scheme^^>\cite{sscp} to explore the interplay between competence and 
disseminating dynamics in the more realistic case of epidemiological parameters 
dependent on the competence level (see Appendix^^>\ref{appendix:B}). Lastly, we 
investigate how the fake news' 
diffusion would impact differently on different classes of the population 
defined in terms of their capabilities of interacting with information.

\subsection{Test 1: Numerical quasi-invariant limit}\label{test:1}

In this test we show that the mean-field Fokker-Planck 
system^^>\eqref{eq:fokkerplanck4}--\eqref{eq:fokkerplanck7} obtained under the 
quasi-invariant scaling^^>\eqref{eq:quasiinvariant1} 
and^^>\eqref{eq:quasiinvariant2} is a good approximation of the Boltzmann 
models^^>\eqref{eq:seiscompetenza1} when $\epsilon \ll 1$. We do so by using a 
Monte Carlo method with $N = 10^4$ particles, starting with a uniform 
distribution of competence $f_0(x) = \frac12\chi(x\in[0,2])$, where 
$\chi(\cdot)$ is the indicator function, and performing various iterations 
until the stationary state was reached; next, the distributions were averaged 
over the next 500 iterations.
We considered constant competence-related parameters $\lambda_{CJ} = 
\lambda_{BJ}$ and $\lambda_J = \lambda_{CJ} + \lambda_{BJ}$ as well as a 
constant variance $\sigma$ for the random variables $\eta_{HJ}$. 

\begin{figure}
\includegraphics[width=0.5\columnwidth]{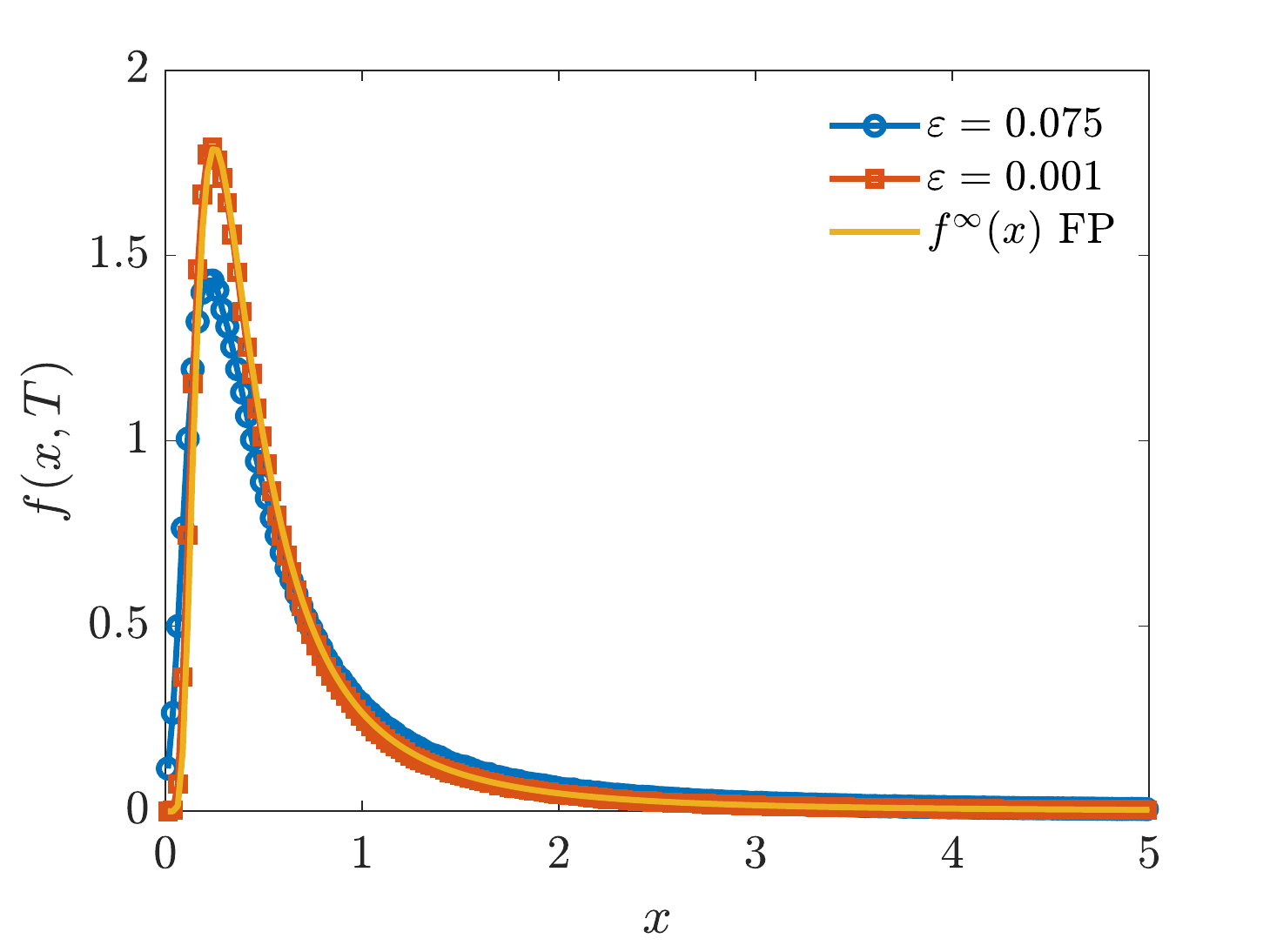}%
\hskip -.1cm
\includegraphics[width=0.5\columnwidth]{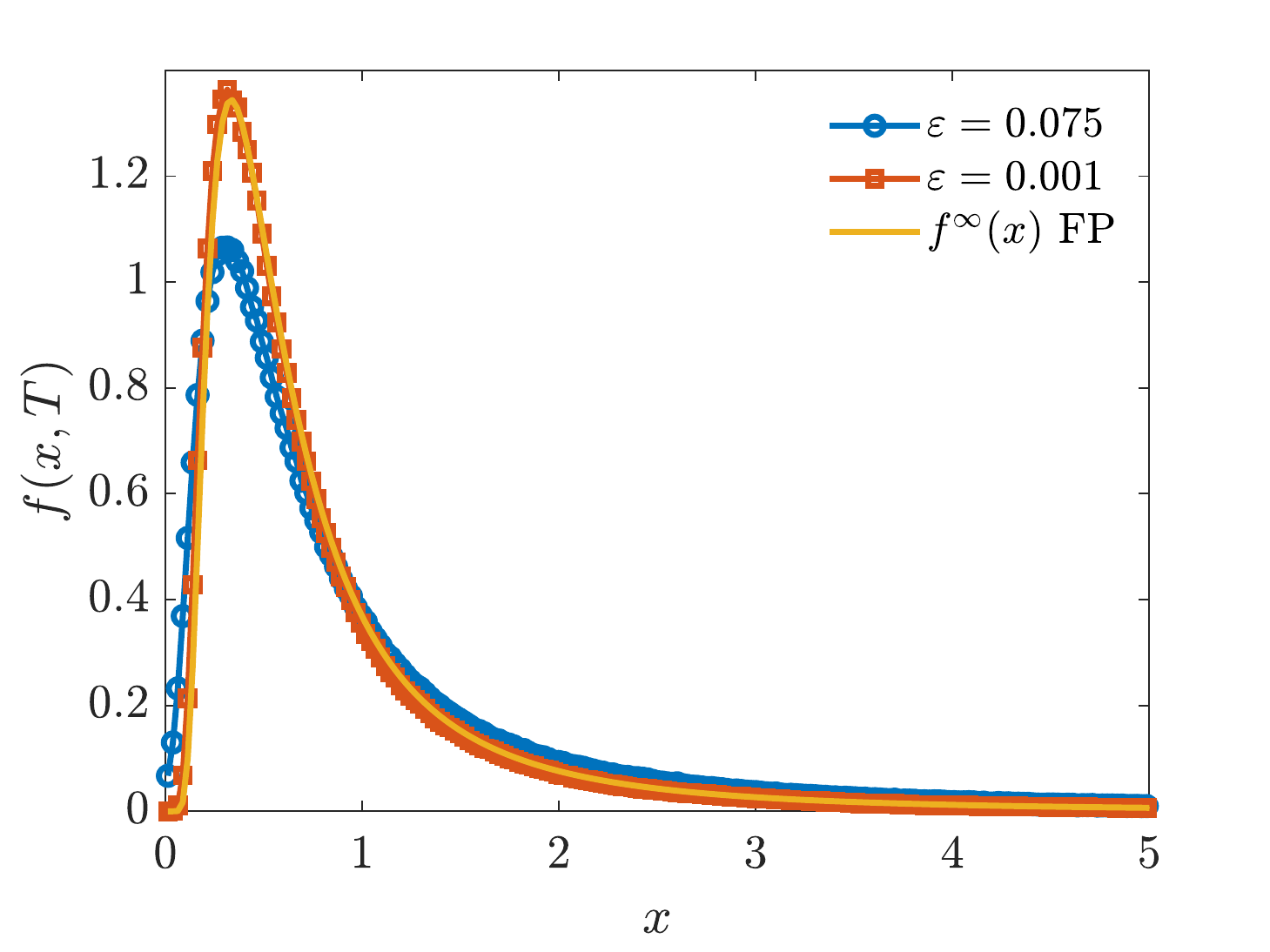}
\caption{Test 1. Comparison of the competence distributions at the 
end of epidemics for system^^>\eqref{eq:seiscompetenza1} with the explicit 
Fokker-Planck solution^^>\eqref{eq:seirFPfinal} with scaling parameters 
$\epsilon=0.075,0.001$. We considered the case  $m_B = 0.75$ (left) and $m_B = 
1$ (right).}
\label{fig:FPvalidation}
\end{figure}

In Figure^^>\ref{fig:FPvalidation}, we plotted the results for 
$(\lambda,\sigma) = (0.075,0.150)$ (circle-solid, teal) and for 
$(\lambda,\sigma) = (0.001,0.002)$ (square-solid, ochre): those choices 
correspond to a scaling regime of $\epsilon=0.075$ and $\epsilon =0.001$, 
respectively, with $\mu = 2$. 
Finally, we assumed that $m_B = 0.75$ (left) and $m_B = 1$ (right).

Directly comparing the Boltzmann dynamics equilibrium with the explicit 
analytic solution of the Fokker-Planck regime shows that if $\epsilon$ is small 
enough, Fokker-Planck asymptotics provide a consistent approximation of the 
steady states of the kinetic distributions.

\subsection{Test 2: Learning dynamics and fake news 
dissemination}\label{test:2}
For this test, we applied the structure-preserving scheme to 
system \eqref{eq:fokkerplanck4}--\eqref{eq:fokkerplanck7} in a more realistic 
scenario featuring an interaction term  dependent on the competence level of 
the agents, as well as a competence-dependent delay during which agents 
evaluate the information and decide how to act.
In this setting, we refer to the recent Survey of Adult 
Skills (SAS) made by the OECD^^>\cite{PIAAC19}: in particular, we focus on 
competence understood as a set of information-processing skills, especially 
through the lens of literacy, defined^^>\cite{PIAAC19} as \lq\lq the ability to 
understand, evaluate, use and engage with written texts in order to participate 
in society\rq\rq. One of the peculiarities that makes the SAS, which is an 
international, multiple-year spanning effort in the framework of the PIAAC 
(Programme for the International Assessment of Adult Competencies) by the 
OECD, interesting in our case is that it was administered digitally to more than 
70\% of the respondents. Digital devices are arguably the most important 
vehicle for information diffusion in OECD countries, so that helps to keep 
consistency. 

Literacy proficiency was defined through $6$ increasing levels; 
we therefore consider a population partitioned in $6$ classes based on the 
competence level of their occupants, equated to the score of the literacy 
proficiency test of the SAS, normalized. Thus, we chose a log-normal-like 
distribution
\[
f(x) = \frac{1}{(\tilde\xi - x)\tilde\sigma\sqrt{2\pi}}
       \cdot \e^{- \frac{(\log(\tilde\xi-x)-\tilde\mu)^2}{2\tilde\sigma^2}},
\]
where $\tilde\xi=5$, $\tilde \mu \approx 0.85$ and $\tilde \sigma \approx 0.22$ 
to make $f(x)$ 
agree with the empirical findings in^^>\cite{PIAAC19}. The computational domain is restricted
to $x\in[0,5]$ and stationary boundary conditions have been applied as 
described in Appendix^^>\ref{appendix:B}.

Initial distributions for the epidemiological compartments were set as
\[%\label{key}
f_S(x,0) = \rho_S f(x), \quad f_E(x,0) = \rho_E f(x), \quad f_I(x,0) = \rho_I f(x),\quad
f_R(x,0) = \rho_R f(x), 
\]
with $\rho_I = 10^{-2}$, $\rho_S= 1 - \rho_I$ and $\rho_E = \rho_R = 0$.

The contact rate $\beta(x,x_*)$ was set as
\begin{equation}\label{eq:beta}
\beta(x,x_*) = \frac{\beta_0}{(1 + x^2)(1 + x_*^2)}\chi(\abs{x - x_*}\le \Delta)
\end{equation}
with $\Delta =2$ 
on the hypothesis that interactions occur more frequently among people with a 
similar competence level and are higher for people with lower competence levels.

The rate $\delta$ at which the information is evaluated by the agents, who 
therefore exits the exposed class, was set to be 
\begin{equation}
\delta(x) = \delta_R+(\delta_L-\delta_R) \frac1{1 + \e^{a(b - x)}},
\end{equation} 
with $\delta_L = 1$, $\delta_R=5$, $a = 2$ and $b=2.5$. Here, 
we are taking into 
account that people with higher efficacy at identifying fake news spend  
significantly more time on conducting their evaluations than people with lower 
efficacy^^>\cite{LEEDER2019100967}. In this specific test case the time range for the
evaluation of the information spans
between 1 day and about 5 hours. The values were purposely chosen rather 
large compared to realistic values in order to highlight also the behavior of the exposed 
compartment. Finally, we set $\gamma = 0.2$, which correspond to an average fake news duration of 5 days,
and $\alpha = 0.2$, so that individuals have a moderate possibility to remember the
fake news and become immune to it, and assume $\eta=0.1$,
namely in this test we do not relate the decision to spread or not fake news to 
the level of competence.

We investigate the relation between the 
dissemination-related component of the model and the competence-related one, 
which entails that agents can learn, i.e., increase their competence level, 
both from the background and from direct binary interactions. Under the 
assumption that $\lambda_{CJ} + \lambda_{BJ} = \lambda_J$, which is a 
conservative choice: the expected value of the competence gained through 
interactions cancels out the one lost due to forgetfulness, in this latter 
process two main parameters are involved: $\lambda = \lambda_J$ (i.e., all 
compartments have the same learning rate) and $m_B$, 
which is the mean of the background competence variable^^>$z$. 

\begin{figure}
\centering
\includegraphics[width=0.5\columnwidth]{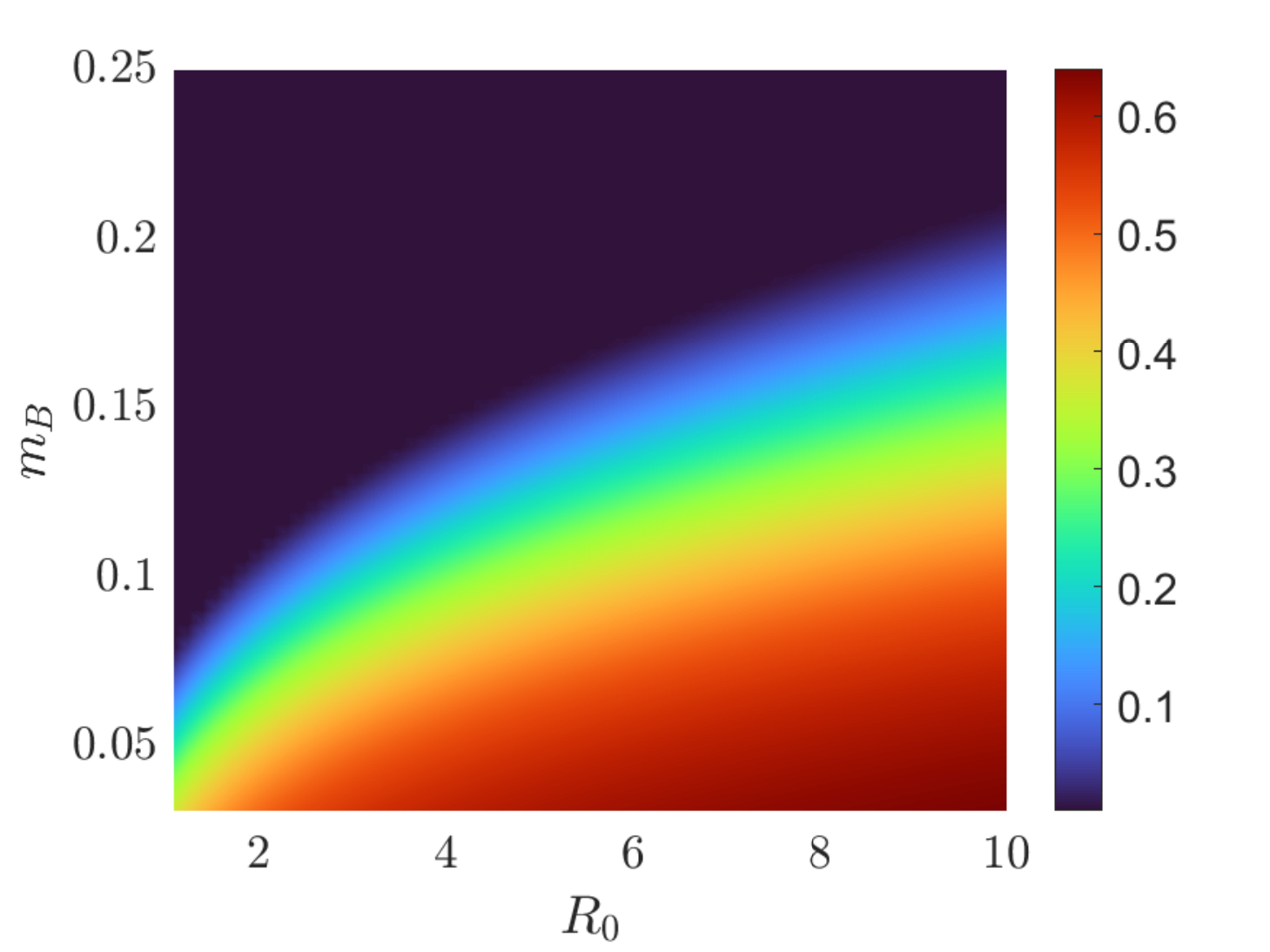}\hskip
 -.1cm
\includegraphics[width=0.5\columnwidth]{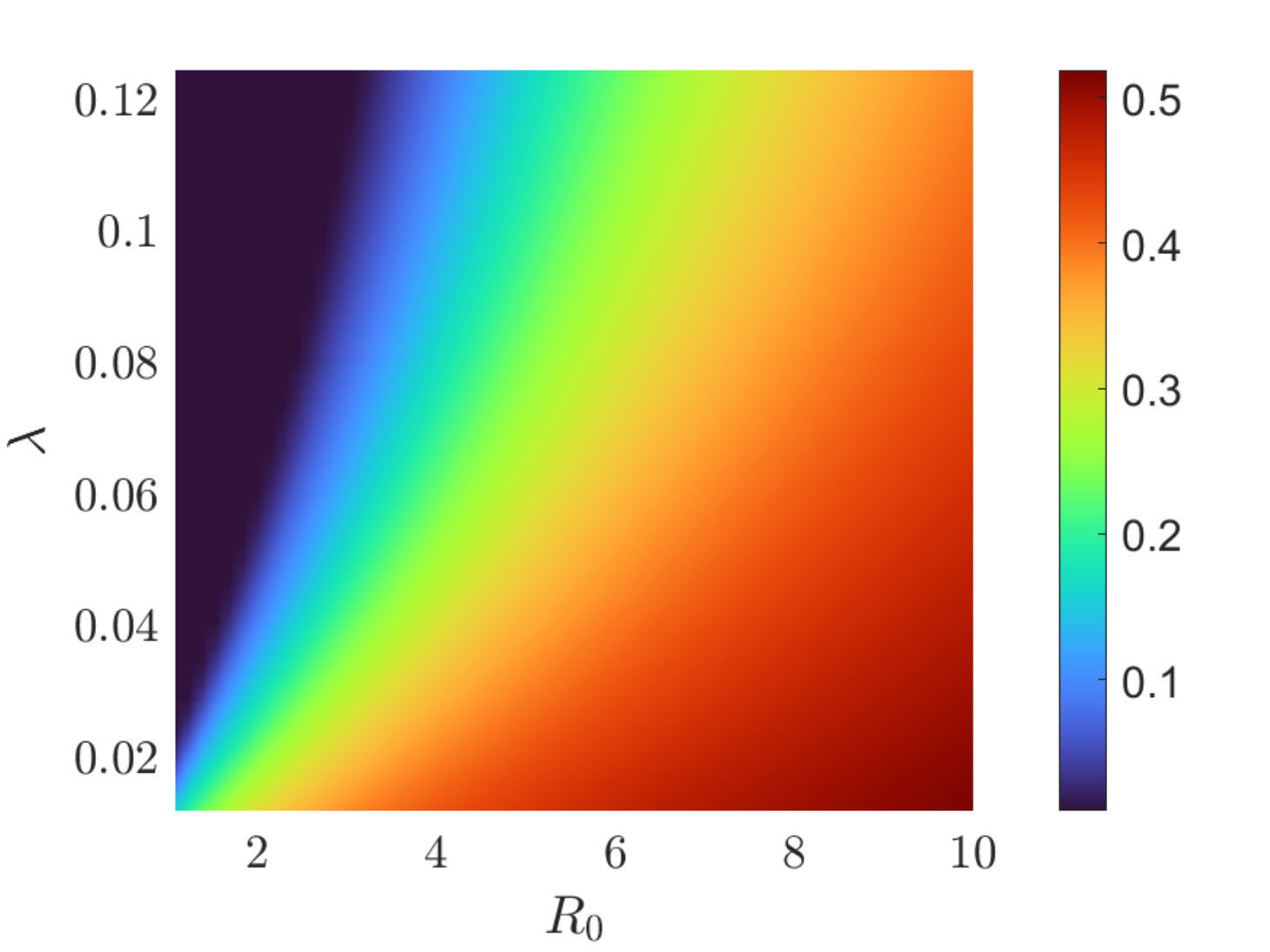}
\caption{Test 2. Interplay between competence levels and dissemination 
dynamics. Contour plots of the highest number of infected in relation to the  
reproduction number^^>\eqref{eq:reproductionnumber} $R_0 \in [1.1,10]$ and the 
background competence mean $m_B \in [0.03125, 0.25]$ (left) and learning rate 
$\lambda \in [0.0125, 0.125]$ (right).}
\label{fig:interplay}
\end{figure}

For what 
concerns the dissemination-related component, instead, the main factor is the 
reproduction number $R_0$^^>\eqref{eq:reproductionnumber}. Hence, we measured 
the differences on the spread of fake news varying these three parameters. In 
Figure^^>\ref{fig:interplay} (left) we show the highest portion of spreaders in 
relation to the mean of background^^>$m_B$ and to the reproduction 
number^^>$R_0$; in the right image $\lambda$ is opposed to $R_0$.

To perform the test, we leveraged the structure-preserving numerical 
scheme^^>\cite{sscp} whose details are presented for 
convenience in Appendix^^>\ref{appendix:B}.
In both images of Figure^^>\ref{fig:interplay} we see transition effects: the 
learning process triggered by the competence dynamics is capable of slowing 
down the dissemination of fake news in the population, even to the point of 
preventing it to take place. In the first case, the mean of the background 
$m_B$, i.e., the mean of the distribution of the background competence variable 
$z$, which we assumed uniformly distributed, varies between $0.03125$ and 
$0.25$, while the reproduction number $R_0$ varies between $1.1$ and $10$. In 
the second case, we left untouched $R_0$, while $\lambda$ varies between 
$0.0125$ and $0.125$ with a background mean $m_B = 0.125$. We can see that the 
mean of the background has a more pronounced impact on the slowing of the 
diffusion of fake news, with a steeper transition effect.

\subsection{Test 3: Impact of the different competence levels}
In this final test we considered how much of an impact the competence level can 
have on the dissemination of fake news in the population. We simulated the 
mean-field model^^>\eqref{eq:fokkerplanck4}--\eqref{eq:fokkerplanck7}  assuming the same 
competence-dependent contact rate $\beta(x,x_*)$ of Test 2, in 
this case with $\beta_0=4$, as well the same delay rate $\delta(x)$ and the 
same $\gamma$, but we additionally assume that the decision to spread or not a
fake news is affected by the level of competence. This is somewhat controversial
in the literature since other factors also affect this behavior like the age
of individuals (tests carried out on young people have shown independence from the
competence in the decision to share a fake news in contrast to what happens in
in older people, see \cite{PIAAC19,LEEDER2019100967}). To emphasize this effect we assume
\begin{equation}
\eta(x)=1-e^{-kx^2},
\end{equation}  
with $k=0.1$. Thus individuals with high level of competence rarely decide to spread
fake news.

\begin{figure}
\centering
\includegraphics[width=0.495\columnwidth]{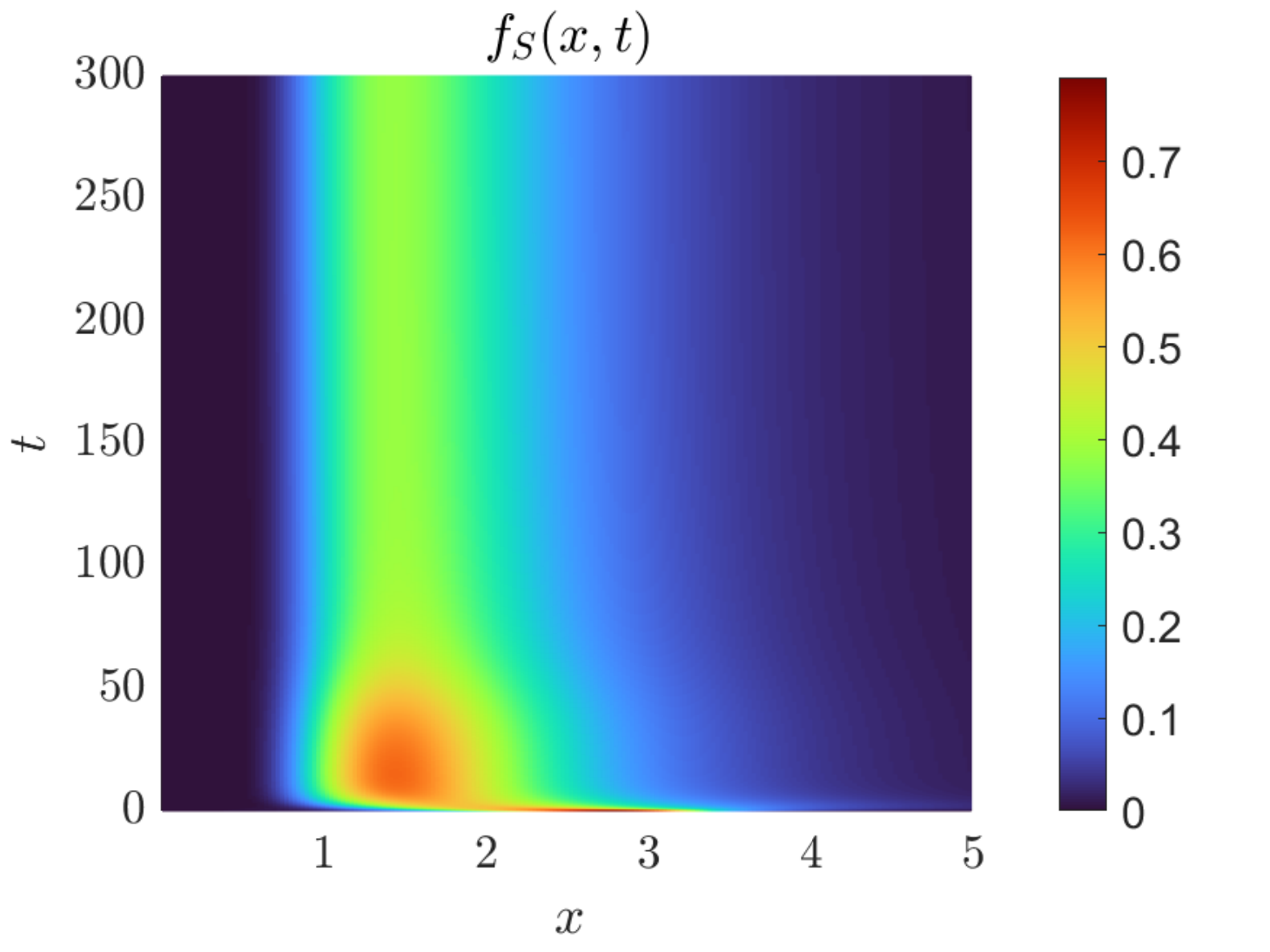}
\includegraphics[width=0.495\columnwidth]{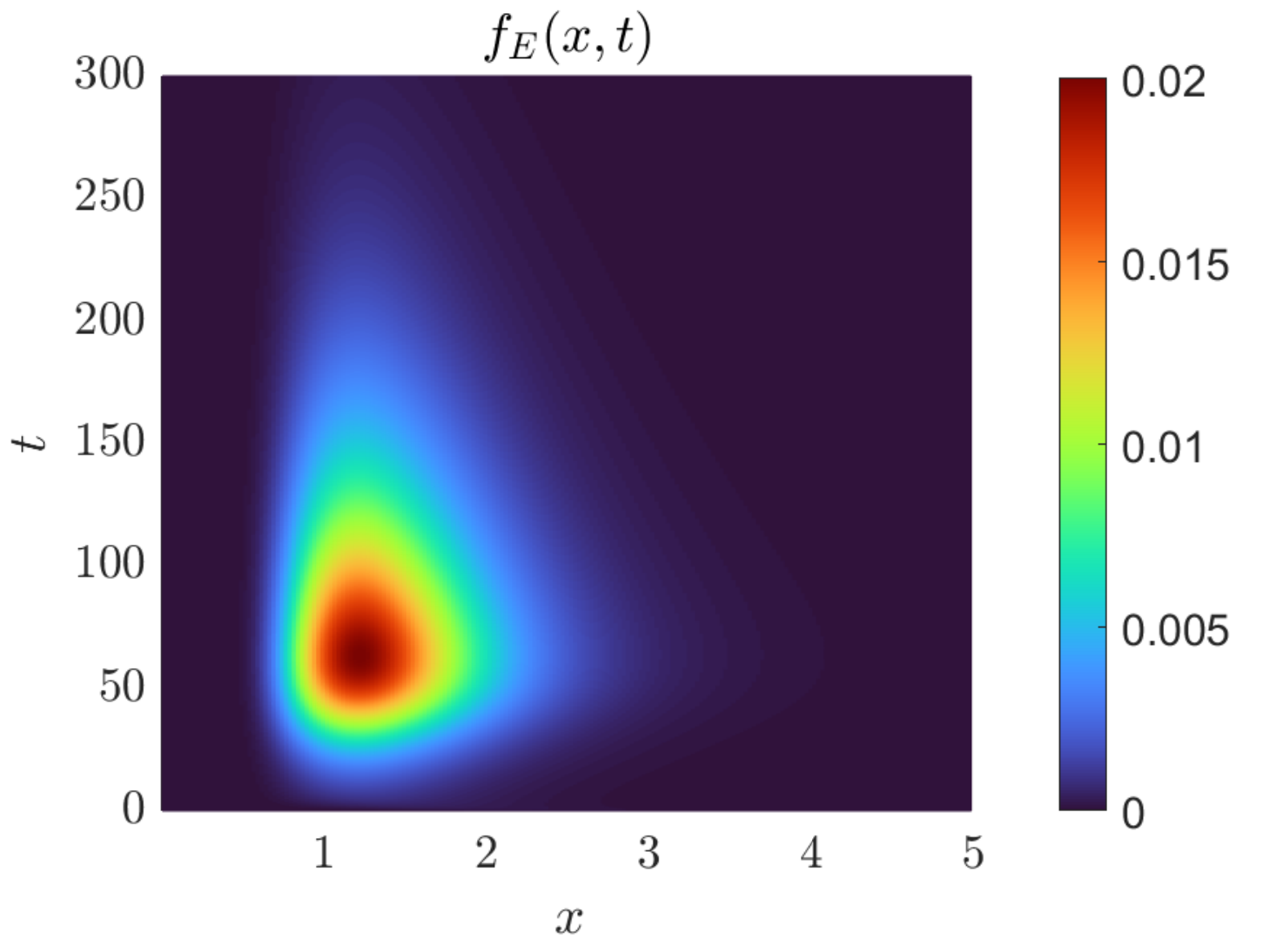}
\includegraphics[width=0.495\columnwidth]{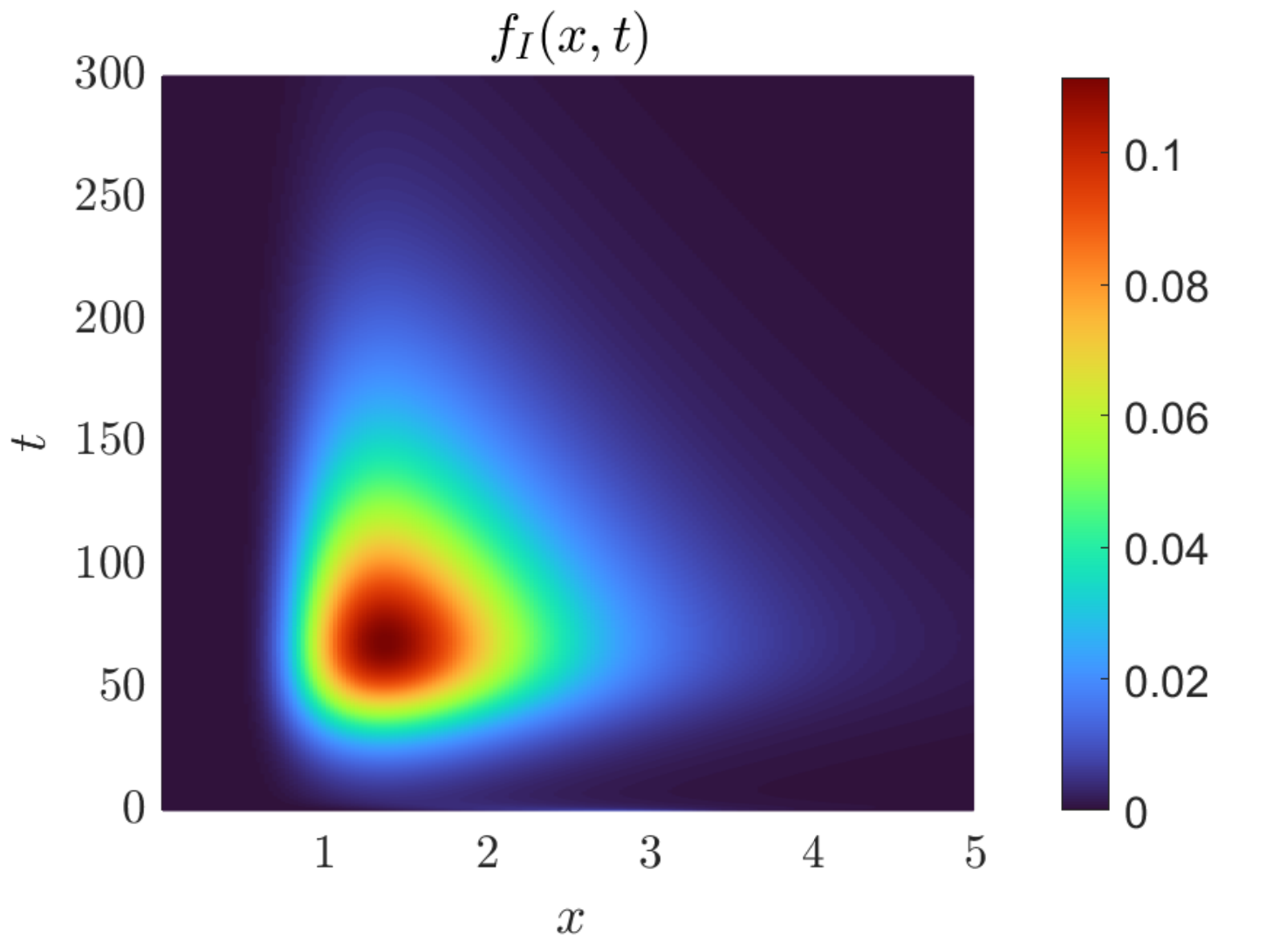}
\includegraphics[width=0.495\columnwidth]{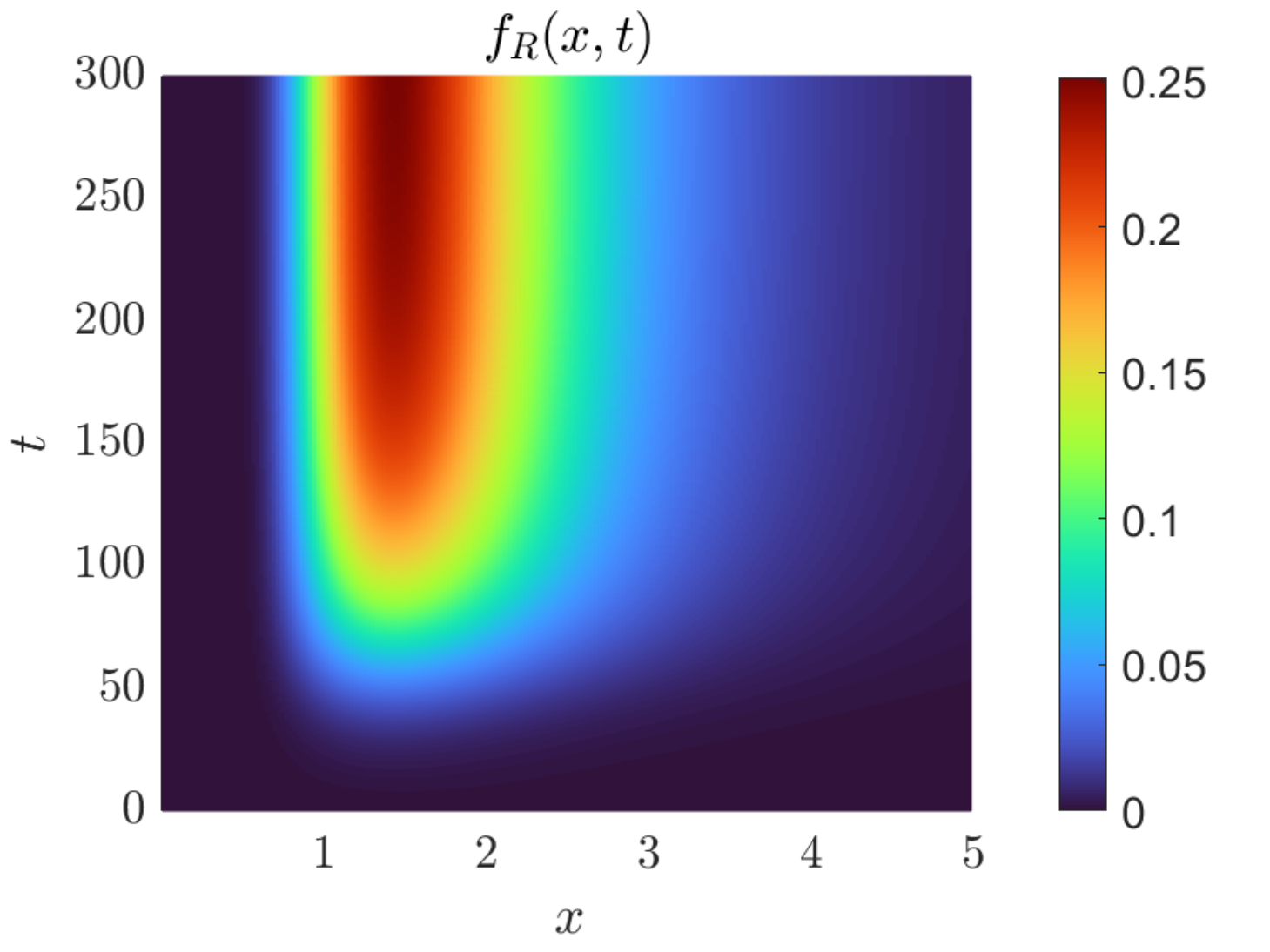}
\caption{Test 3. Time evolution of the competence distribution for the kinetic 
model^^>\eqref{eq:seiscompetenza1} with competence parameters $\lambda_{BJ} = 
\lambda_{CJ} = 0.125$, $\lambda_J= \lambda_{BJ} + \lambda_{CJ}$, with a 
background competence mean of $m_B$ = 0.125. We considered $\alpha=0.1$, 
$\beta_0 = 4$, $\gamma =0.2$. Top left: susceptible; top right: exposed; bottom 
left: infected; bottom right: removed.}\label{fig:seir-piaac-evolution-01}
\end{figure}

\begin{figure}
\centering
\includegraphics[width=0.495\columnwidth]{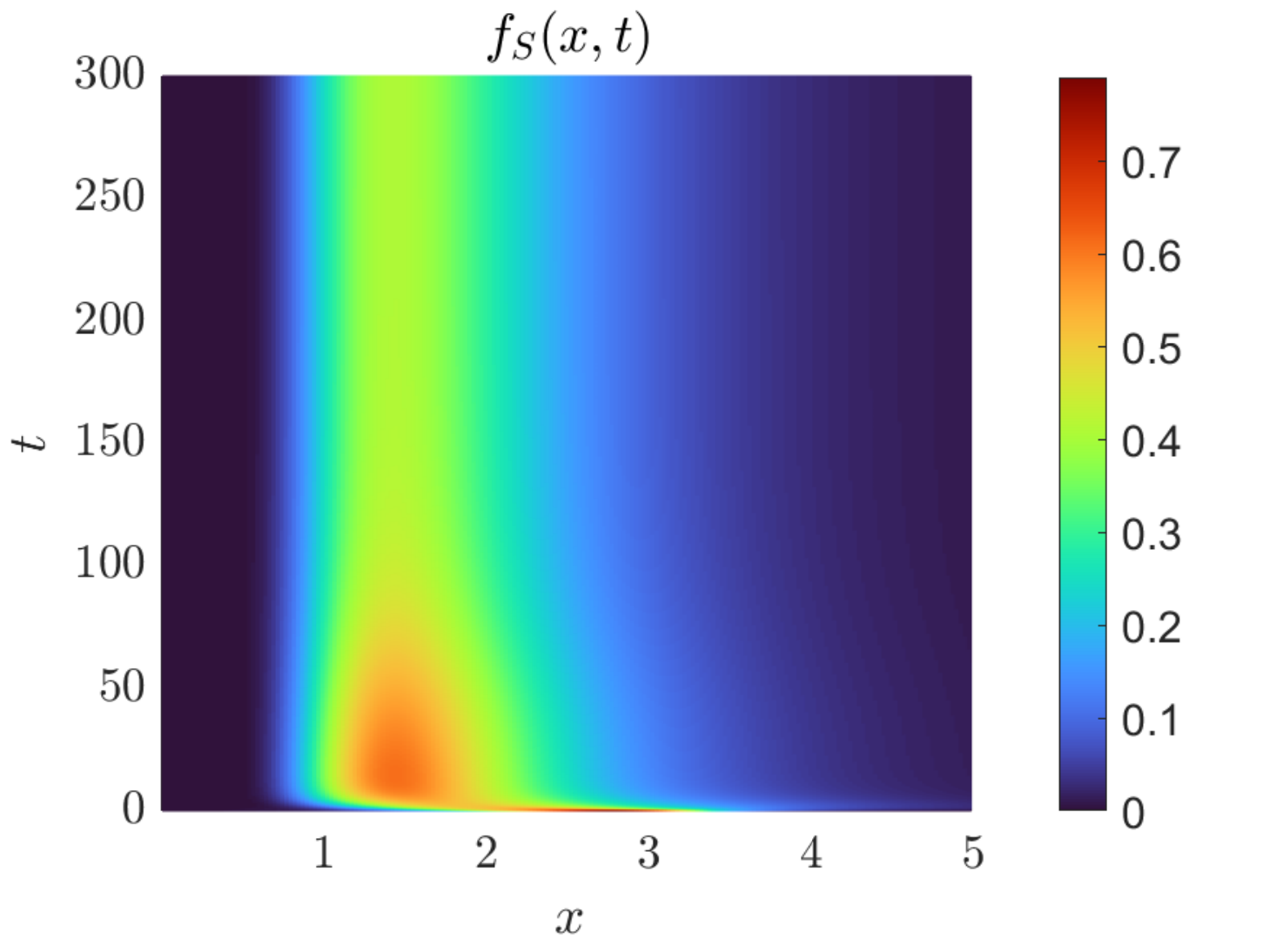}
\includegraphics[width=0.495\columnwidth]{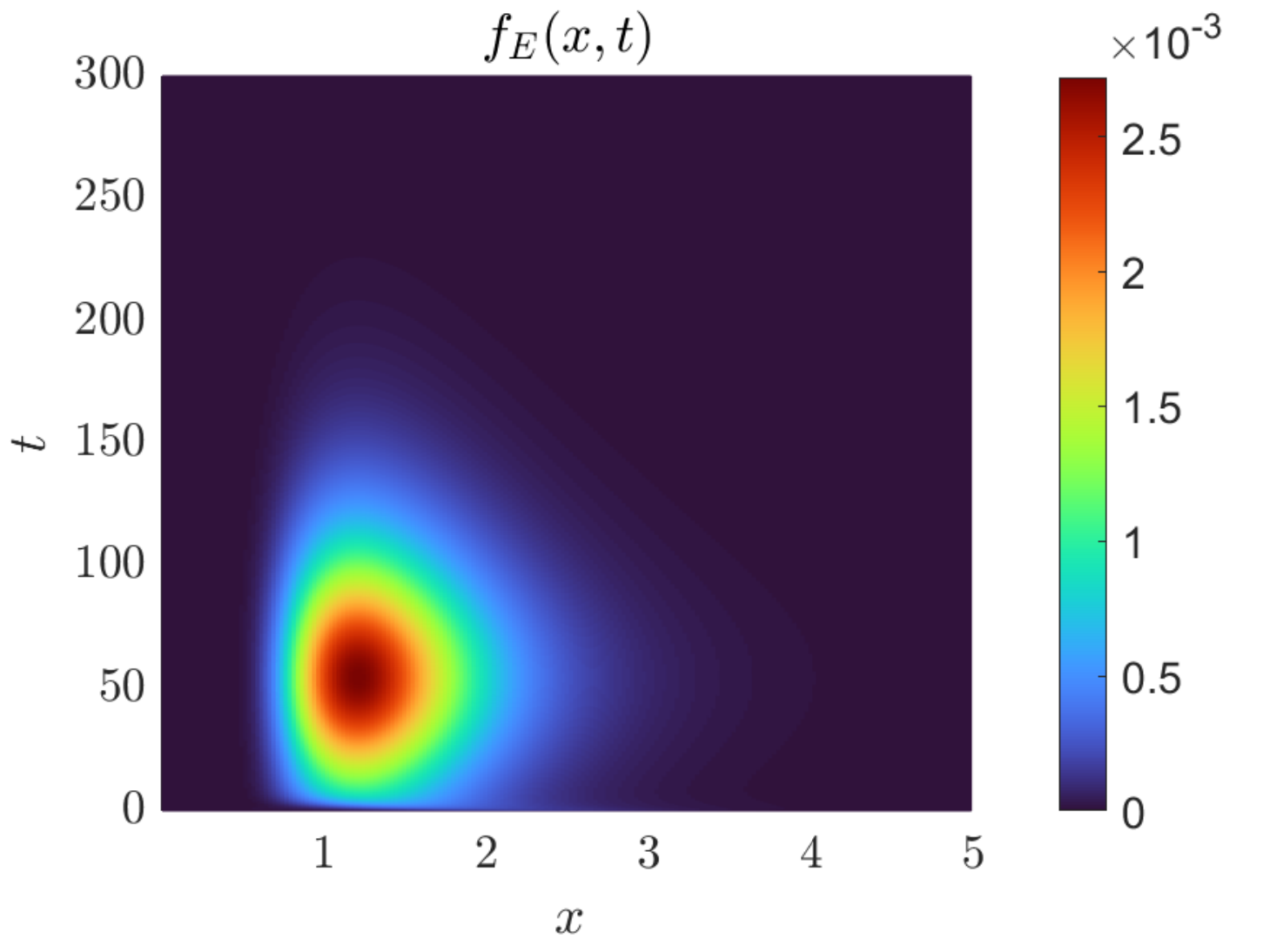}
\includegraphics[width=0.495\columnwidth]{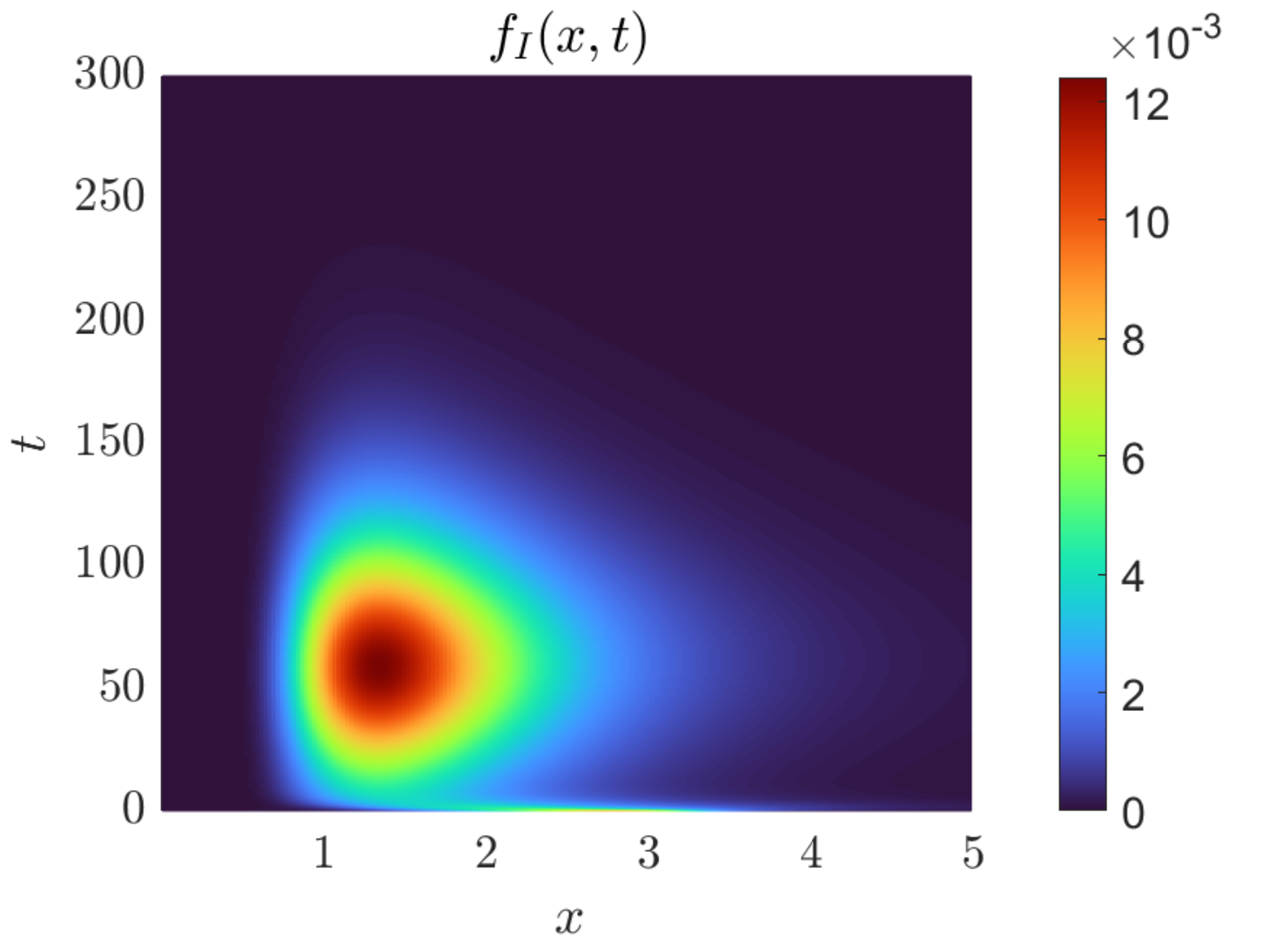}
\includegraphics[width=0.495\columnwidth]{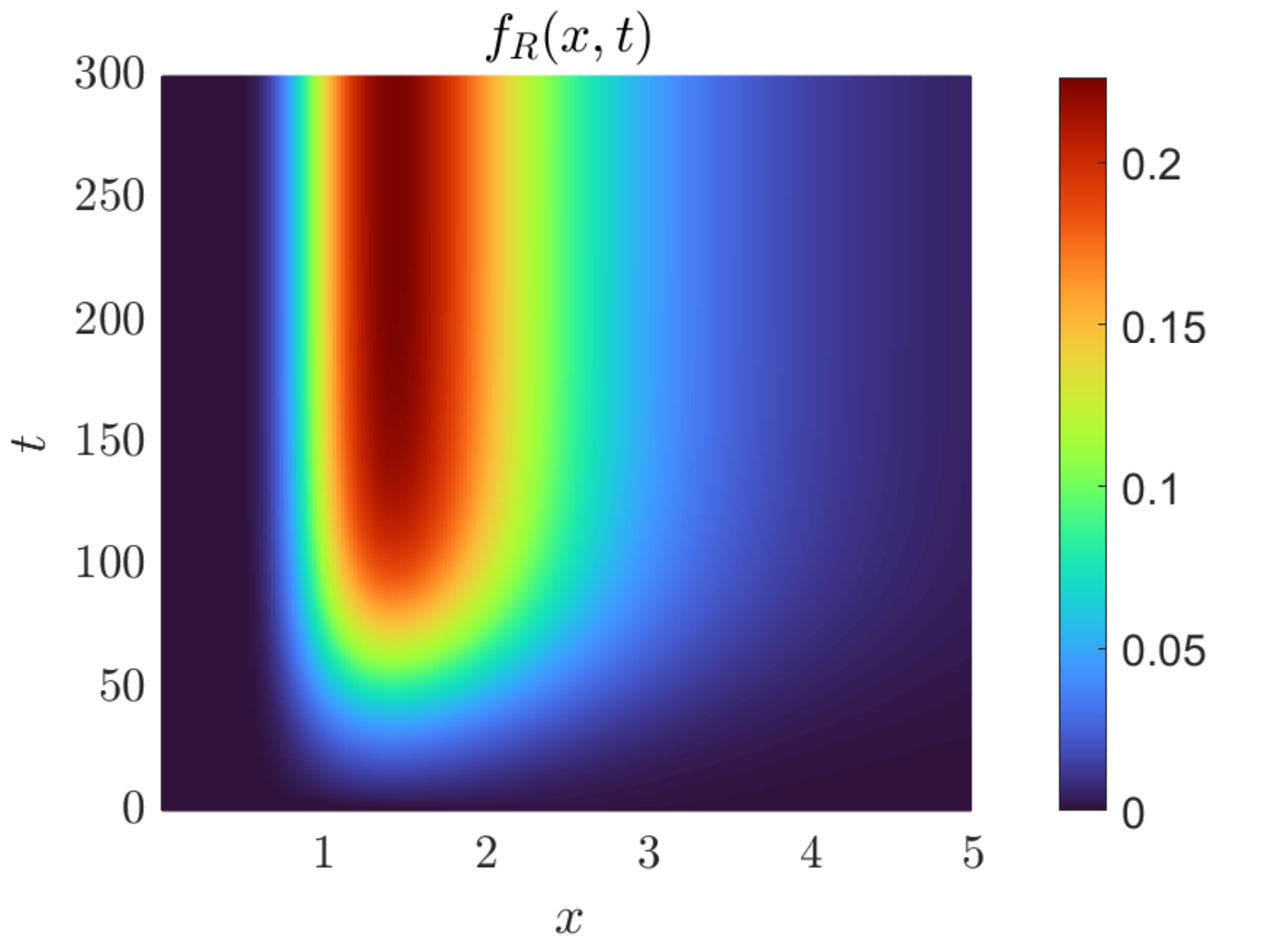}
\caption{Test 3. Time evolution of the competence distribution for the kinetic 
model^^>\eqref{eq:seiscompetenza1} with competence parameters $\lambda_{BJ} = 
\lambda_{CJ} = 0.125$, $\lambda_J= \lambda_{BJ} + \lambda_{CJ}$, with a 
background competence mean of $m_B$ = 0.125. We considered $\alpha=0.9$, 
$\beta_0 = 4$, $\gamma =0.2$. Top left: susceptible; top right: exposed; bottom 
left: infected; bottom right: removed.}\label{fig:seir-piaac-evolution-09}
\end{figure}

In Figure^^>\ref{fig:seir-piaac-evolution-01} we 
report the time evolution of the distributions of susceptible (top left), 
exposed (top right), infected (bottom left) and removed (bottom right) agents 
with competence parameters of $\lambda_{BJ} = \lambda_{CJ} = 0.125$, 
$\lambda_J= \lambda_{BJ} + \lambda_{CJ}$ and $m_B = 0.125$, in the case $\alpha 
= 0.1$. In Figure^^>\ref{fig:seir-piaac-evolution-09}, instead, we show the 
evolution with the same parameters except for a larger probability $\alpha=0.9$ 
of remembering the fake news.

\begin{figure}
\centering\includegraphics[width=0.495\columnwidth]{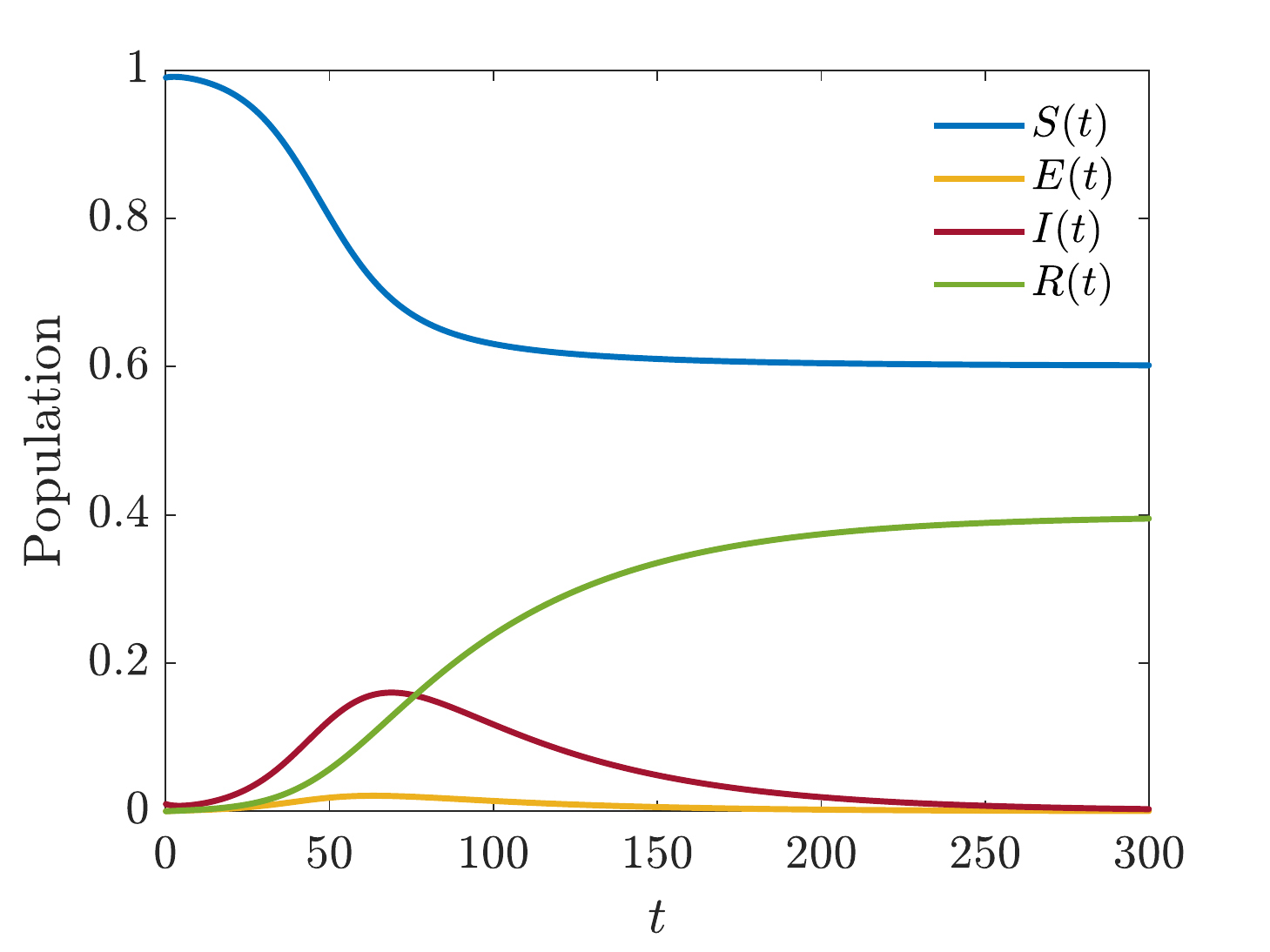}
\centering\includegraphics[width=0.495\columnwidth]{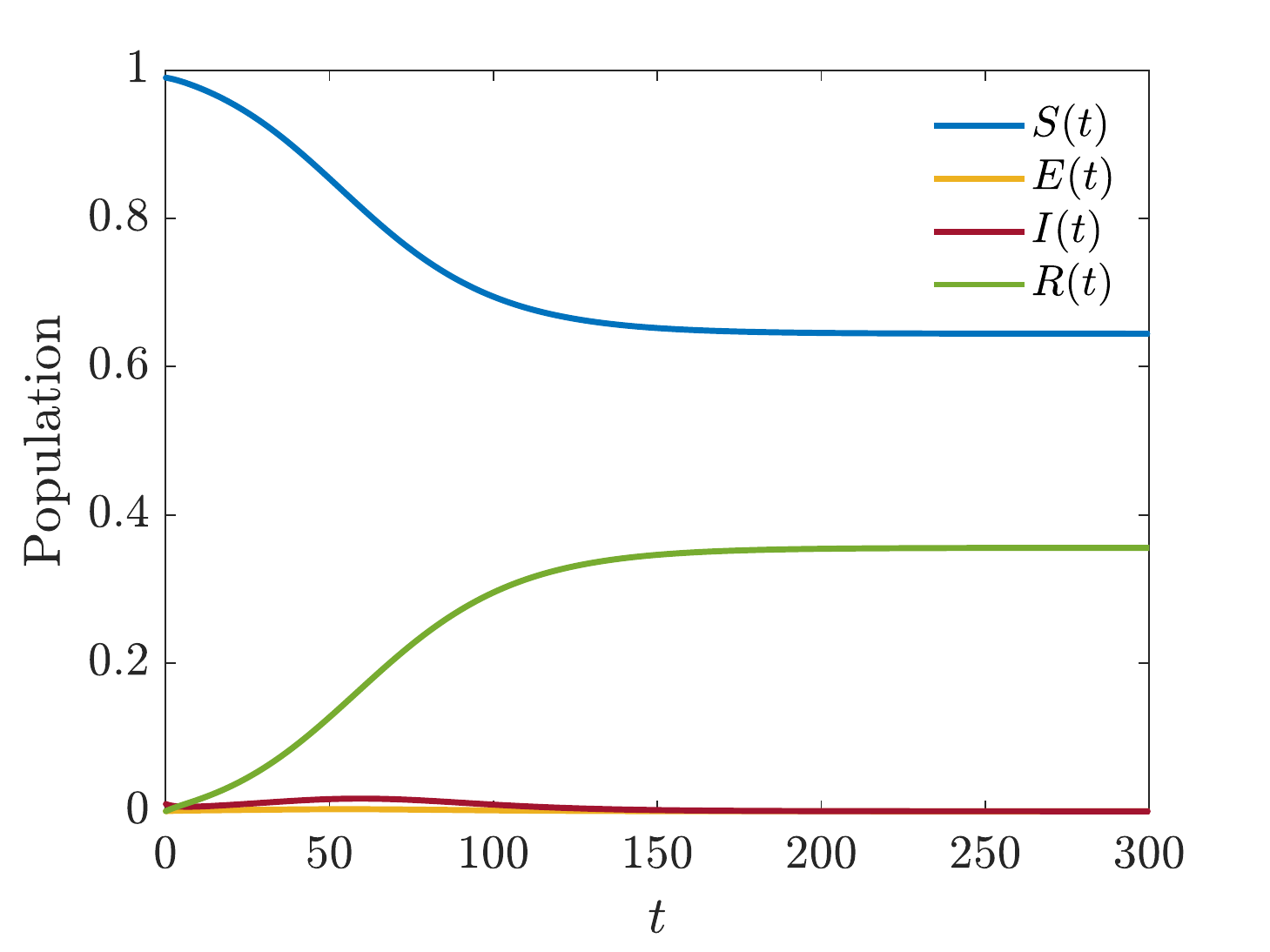}
\caption{Test 3. Evolution in time of the densities for susceptible, exposed, 
infectious and removed agents. Left: $\alpha = 0.1$; right: $\alpha = 0.9$.}
\label{fig:seir-piaac-relative-numbers}
\end{figure}

In Figure^^>\ref{fig:seir-piaac-relative-numbers} are shown the relative 
numbers of susceptible, exposed, infected and removed agents, on the left for 
$\alpha = 0.1$ and on the right for $\alpha = 0.9$.
To measure the effects of the competence, we considered the curve of the 
infected agents depending on their levels accordingly to^^>\cite{PIAAC19} for
$x\in [0,5]$:
\begin{itemize}
\item below level 1: scoring less then 175/500, ($x < 1.75$);
\item level 1: scoring between 176/500 and 225/500, ($x > 1.75$ and $x < 2.25$);
\item level 2: scoring between 226/500 and 275/500, ($x > 2.25$ and $x < 2.75$);
\item level 3: scoring between 276/500 and 325/500, ($x > 2.75$ and $x < 3.25$);
\item level 4: scoring between 326/500 and 375/500, ($x > 3.25$ and $x < 3.75$);
\item level 5: scoring more than 375/500, ($x > 3.75$).
\end{itemize}
\begin{figure}
\centering
\includegraphics[width=0.495\columnwidth]{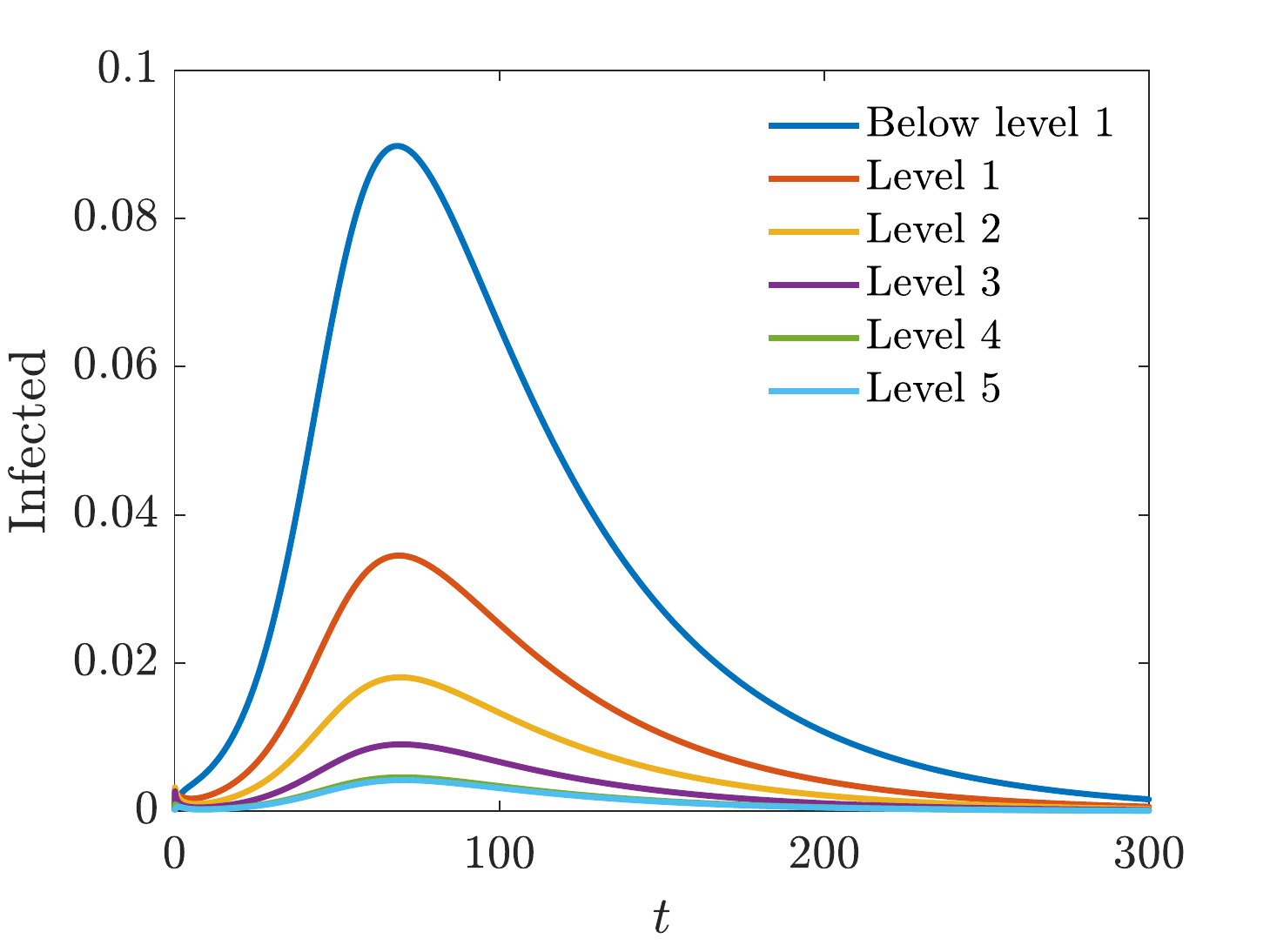}
\includegraphics[width=0.495\columnwidth]{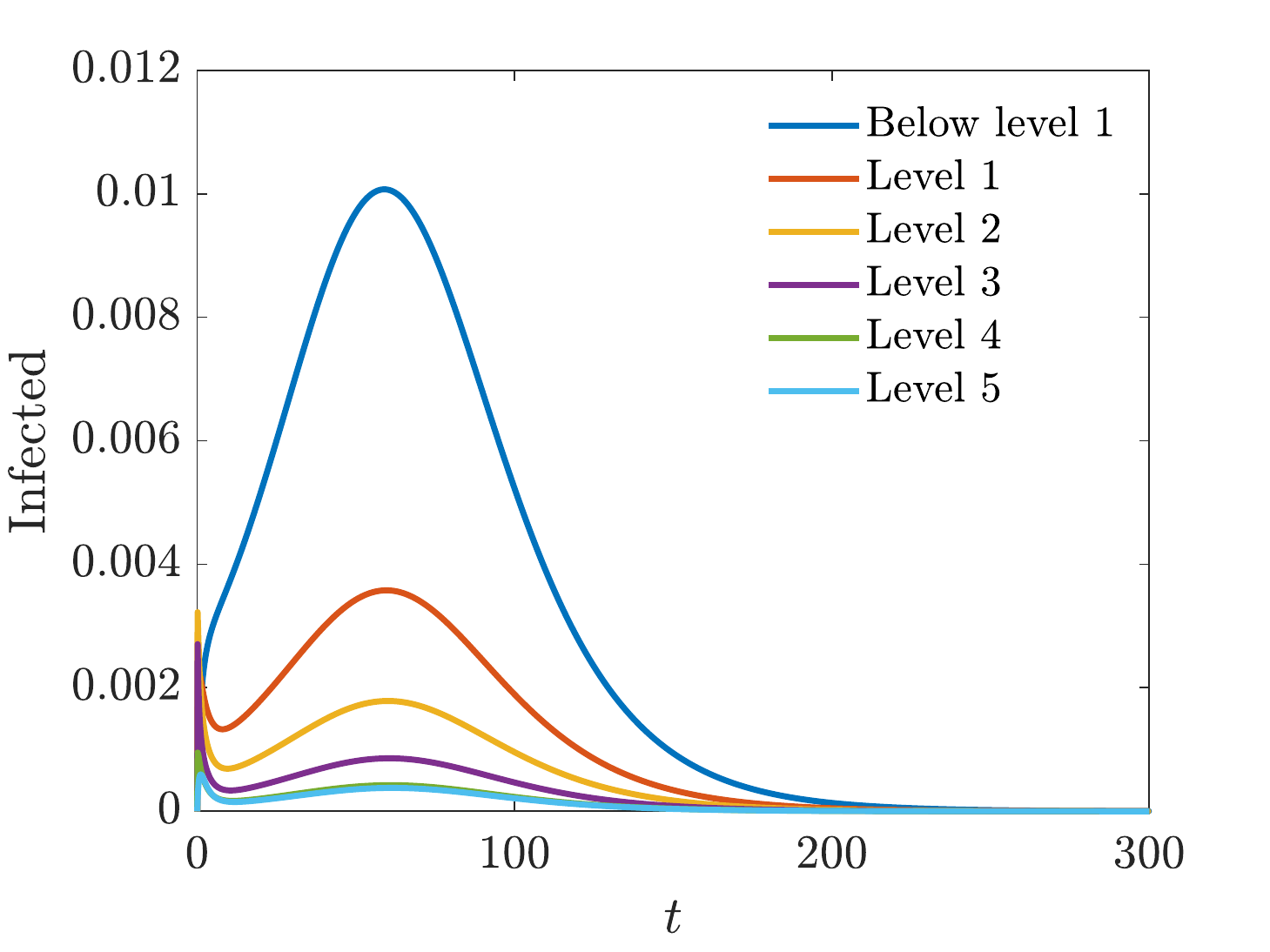}
\caption{Test 3. Evolution of the fraction of infectious agents with different 
levels of competence as defined in^^>\cite{PIAAC19}. Left: $\alpha = 0.1$; 
right: $\alpha = 0.9$.}
\label{fig:seir-piaac-levels}
\end{figure}
Figure^^>\ref{fig:seir-piaac-levels} shows clearly that the more 
competent the individual, the lesser they contribute to the spread of fake 
news, in perfect agreement with the transition effects observed in 
Test^^>\ref{test:2} due to the interplay between competence and the 
dissemination dynamics.
Moreover, we can see how the probability $\alpha$ of detecting fake news 
influences its dissemination in the population: a lower probability implies a 
higher peak of infected agents for each competence level, as well as a slower 
spread overall.

\section{Concluding remarks}\label{sec:conclusions}
In this paper, we introduced a compartmental model for fake news dissemination 
that also considers the competence of individuals. In the model, the concept of 
competence is not introduced as a static feature of the dynamic, but as an 
evolutionary component that takes into account both learning through 
interactions between agents and interventions aimed at educating individuals in 
the ability to detect fake news.

From a mathematical viewpoint the competence dynamics has been introduced as a 
Boltzmann interaction term in the corresponding system of differential 
equations. A suitable scaling limit, permits to recover the corresponding 
Fokker-Planck models and then the resulting stationary states in terms of 
competence. These, in agreement with \cite{PARESCHI2017201}, are given 
explicitly by Gamma distributions. 

The numerical results demonstrate the model's ability to correctly describe the 
interplay between fake news dissemination and individuals' level of competence, 
highlighting transition phenomena at the level of expertise that allow fake 
news 
to spread more rapidly.

Future developments of the model will be considered in particular in the case 
of networks, in order to describe the spread of fake news on social networks 
and present plausible scenarios useful to limit the spread of false 
information. This can be done by following an approach similar to that of 
kinetic models for opinion formation on networks^^>\cite{network1}. 
Another challenging aspect concerns the matching of the model with realistic 
data that requires the introduction of quantitative aspects not always easy to 
identify^^>\cite{maleki2021using,datarepository}. One of the main 
applications will be related to combating misinformation in the vaccination campaign
against COVID-19.

\section*{Acknowledgments}
This work was partially supported by MIUR (Ministero dell’Istruzione, 
dell’Università e della
Ricerca) PRIN 2017, project “\textit{Innovative numerical methods for 
evolutionary partial differential equations and applications}”, code 2017KKJP4X.

\appendix
\section{Proof of Theorem^^>\ref{teo:fouriermetric}}\label{appendix:A}
We provide here a proof for Theorem^^>\ref{teo:fouriermetric}. The proof is 
identical to^^>\cite{PhysRevE.102.022303}; we develop it here, too, for 
completeness, only in the case $\alpha = \eta = 0$. If \mbox{$H \in \ins{S, E, 
I}$} we 
have
\begin{equation}\label{eq:defQ+}
\begin{aligned}
\sum_{\mathclap{J \in \ins{S,E,I}}}
    \ft Q(\ft f_H, \ft f_J)(\xi, t) &=
\sum_{\mathclap{J \in \ins{S,E,I}}}
    \expvalue{\ft f_H(A_{HJ} \xi -\lambda_{BH}z, t)}\ft f_J(\lambda_{CJ}\xi,t)
                                    - 
\sum_{\mathclap{J \in \ins{S,E,I}}}
    \ft f_H(\xi,t) J(t)\\
                                    &=
\ft Q_+(\ft f_H)(\xi, t) - \ft H(\xi,t)
\end{aligned}
\end{equation}
where the second equality follows from mass conservation. 
In^^>\eqref{eq:defQ+}, we have defined^^>$\ft Q_+(\ft f_H)(\xi, t)$ as
\[
\ft Q_+(\ft f_H)(\xi, t) = \sum_{\mathclap{J \in \ins{S,E,I}}}
    \expvalue{\ft f_H(A_{HJ} \xi -\lambda_{BH}z, t)}\ft f_J(\lambda_{CJ}\xi,t),
\]
where $A_{HJ}$ has been defined originally in^^>\eqref{eq:defAHJ}. Thus, now 
system^^>\eqref{eq:seiscompetenzaFourier} reads
\begin{equation}
\left\lbrace
\begin{aligned}
\pd{\ft f_S(\xi,t)}t + \ft f_S(\xi, t) &= -\beta I(t) \ft f_S(\xi,t)
                                          + \gamma \ft f_I(\xi,t)
                                          + \ft Q_+(\ft f_I)(\xi,t)\\ 
\pd{\ft f_E(\xi,t)}t + \ft f_E(\xi, t) &= \beta I(t) \ft f_S(\xi,t)
                                          - \delta \ft f_E(\xi,t)
                                          + \ft Q_+(\ft f_E)(\xi,t)\\
\pd{\ft f_I(\xi,t)}t + \ft f_I(\xi, t) &= \delta \ft f_E(\xi,t)
                                          - \gamma \ft f_I(\xi,t)
                                          + \ft Q(\ft f_I)(\xi,t),
\end{aligned}
\right.
\label{eq:seiscompetenzaFourier1}
\end{equation}
To ensure positivity of all coefficients on the right-hand side (since 
\mbox{$I(t) < 1$} and^^>\mbox{$\beta, \gamma, \delta < 1$} we can add^^>$\ft 
f_J(\xi,t)$ to each side, where $J$ equals $S$, $E$ and^^>$I$ in the first, 
second and^^>third equation, respectively, to resort to the equivalent system
\begin{equation}
\left\lbrace
\begin{aligned}
\pd{\ft f_S(\xi,t)}t + 2\ft f_S(\xi, t) &= (1 - \beta I(t)) \ft f_S(\xi,t)
                                           + \gamma \ft f_I(\xi,t)
                                           + \ft Q_+(\ft f_I)(\xi,t)\\ 
\pd{\ft f_E(\xi,t)}t + 2\ft f_E(\xi, t) &= \beta I(t) \ft f_S(\xi,t)
                                           (1 - \delta) \ft f_E(\xi,t)
                                           + \ft Q_+(\ft f_E)(\xi,t)\\
\pd{\ft f_I(\xi,t)}t + 2\ft f_I(\xi, t) &= \delta \ft f_E(\xi,t)
                                           (1 - \gamma) \ft f_I(\xi,t)
                                           + \ft Q(\ft f_I)(\xi,t),
\end{aligned}
\right.
\label{eq:seiscompetenzaFourier2}
\end{equation}
in which all coefficients on the right-hand side are positive.

Now, let $\ft f_J(\xi,t)$ and $\ft g(\xi, t)$ be two solutions of the 
system^^>\eqref{eq:seiscompetenzaFourier2}. We look at the time behavior of 
the^^>$d_2$ metric of their difference, where the Fourier-based metric was 
defined in^^>\eqref{eq:defd2}; therefore we define
\[
h_J(\xi, t) = \frac{\ft f_J(\xi,t) - \ft g_J(\xi,t)}{\abs \xi ^2}.
\]
We see that the metric and $h_J$ are related by
\begin{equation}\label{eq:relationd2hJ}
d_2(f_J, g_J) = \norminf{h_J}
\end{equation}
By its definition, the $h_J$ are solutions of
\begin{equation}
\left\lbrace
\begin{aligned}
\pd{\ft h_S(\xi,t)}t + 2\ft h_S(\xi, t) &= (1 - \beta I(t)) \ft h_S(\xi,t)
                                           + \gamma \ft h_I(\xi,t)
                                           + L_+(\ft f_S)(\xi,t)\\ 
\pd{\ft h_E(\xi,t)}t + 2\ft h_E(\xi, t) &= \beta I(t) \ft h_S(\xi,t)
                                           (1 - \delta) \ft h_E(\xi,t)
                                           + L_+(\ft f_E)(\xi,t)\\
\pd{\ft h_I(\xi,t)}t + 2\ft h_I(\xi, t) &= \delta \ft h_E(\xi,t)
                                           + (1 - \gamma) \ft h_I(\xi,t)
                                           + L_+(\ft f_I)(\xi,t),
\end{aligned}
\right.
\label{eq:seiscompetenzaFourier3}
\end{equation}
where, with $H \in \ins{S, E, I, R}$ are defined as
\[
L_+(\ft f_H)(\xi, t) = \frac{\ft Q_+(\ft f_H)(\xi,t) - \ft Q_+(\ft g_H)(\xi,t)}
                            {\abs \xi^2}.
\]
We can rewrite $L_+(\ft f_H)(\xi, t)$ in full:
\begin{equation}\label{eq:sumL+}
L_+(\ft f_H)(\xi, t) = \sum_{\mathclap{J \in \ins{S,E,I,R}}}
                       \frac{\expvalue{
                            \ft f_H(A_{HJ}\xi - \lambda_{BH}z,t)
                            \ft f_J(\lambda_{CJ}\xi,t)
                          - \ft g_H(A_{HJ}\xi - \lambda_{BH}z,t)
                            \ft g_J(\lambda_{CJ}\xi,t)
                                      }}
                            {\abs \xi^2},
\end{equation}
with the expectation^^>$\expvalue \cdot$ put outside for convenience. As shown, 
e.g., in^^>\cite{intermultiagent}, and since $f_J$ and^^>$g_J$ are solution of 
the SEIS system for the masses and the mean values,  we can profitably bound 
the addends in the sum on the right-hand side of^^>\eqref{eq:sumL+} in terms of 
the functions^^>$h_H$ and^^>$h_J$
\[
\begin{multlined}[t][\columnwidth]
\shoveleft{\abs*{\frac{\expvalue{
                                         \ft f_H(A_{HJ}\xi - \lambda_{BH}z,t)
                                         \ft f_J(\lambda_{CJ}\xi,t)
                                         - \ft g_H(A_{HJ}\xi - \lambda_{BH}z,t)
                                         \ft g_J(\lambda_{CJ}\xi,t)
                                        }}
                              {\abs \xi^2}}
          }\\
\shoveleft[1cm]{\le
               \expvalue*{
                          \abs{\ft f_H(A_{HJ}\xi - \lambda_{BH}z,t)}
                          \abs*{\frac{\ft f_J(\lambda_{CJ}\xi,t)
                                       - \ft g_J(\lambda_{CJ}\xi,t)}
                                     {\abs{\lambda_{CJ}\xi}^2}
                               }
                         \lambda_{CJ}^2
                         }
               }\\
          \shoveright{ + \expvalue*{
                          \abs{\ft g_J(\lambda_J\xi,t)}
                          \abs*{\frac{\ft f_H(A_{HJ}\xi - \lambda_{BH}z,t)
                                       - \ft g_H(A_{HJ}\xi - \lambda_{BH}z,t)}
                                     {\abs{A_{HJ}\xi}^2}
                               }
                          A_{HJ}^2
                         }
                     }\\
\shoveleft[1cm]{\le
              H(t)\sup_\xi \abs*{\frac{\ft f_J(\xi,t) - \ft g_J(\xi,t)}
                                      {\abs \xi ^2}}
                           \lambda_{CJ}^2
             + J(t)\sup_\xi \abs*{\frac{\ft f_H(\xi,t) - \ft g_H(\xi,t)}
                                      {\abs \xi ^2}}
                            \expvalue{A_{HJ}^2}
               }\\
\shoveleft[1cm]{=
             \lambda_{CJ}^2 H(t)\norminf{h_J} + 
             \expvalue{A_{HJ}^2}\norminf{h_H}.
               }\\
\end{multlined}
\]
\vskip-\belowdisplayskip\vskip-\jot %compensating the unnecessary 
%line above
\noindent So we obtain that
\[
\norminf{L_+(\ft f_H)(t)} \le H(t) \sum_{\mathclap{J \in \ins{S,E,I}}}
                            \lambda_{CJ}^2\norminf{h_J(t)} +
                            \norminf{h_H(t)}\sum_{\mathclap{J \in 
                            \ins{S,E,I}}}
                            \expvalue{A_{HJ}^2 J(t)}. 
\]
Multiplying both sides of^^>\eqref{eq:seiscompetenzaFourier3} by^^>$\e^{2t}$ we 
have
\begin{equation*}
\left\lbrace
\begin{aligned}
\pd{[h_S(\xi,t)\e^{2t}]}t  &\le (1 - \beta I(t))\norminf{h_I(t)\e^{2t}}
                                           + \gamma\norminf{h_I(t)\e^{2t}}
                                           + \norminf{L_+(\ft f_S)(t)\e^{2t}}\\ 
\pd{[h_E(\xi,t)\e^{2t}]}t  &\le \beta I(t)\norminf{h_I(t)\e^{2t}}
                                           + (1 - \delta)\norminf{h_E(t)\e^{2t}}
                                           + \norminf{L_+(\ft f_E)(t)\e^{2t}}\\
\pd{[h_I(\xi,t)\e^{2t}]}t  &\le \delta \norminf{h_E(t)\e^{2t}}
                                           + (1 - \gamma)\norminf{h_I(t)\e^{2t}}
                                           + \norminf{L_+(\ft f_I)(t)\e^{2t}},
\end{aligned}
\right.
%\label{eq:seiscompetenzaFourier4}
\end{equation*}
If we integrate from $0$ to $t$ and take the suprema we get
\begin{equation*}
\left\lbrace
\begin{aligned}
\norminf{h_S(\xi,t)\e^{2t}}  &\le \norminf{h_S(0)}
                                   + \int_0^t \biggl[
                                         (1 - \beta I(t))\norminf{h_I(t)\e^{2t}}
                                         + \gamma\norminf{h_I(t)\e^{2t}}
                                         + \norminf{L_+(\ft f_S)(t)\e^{2t}}
                                              \biggr]\, \de s\\ 
\norminf{h_E(\xi,t)\e^{2t}}  &\le \norminf{h_E(0)}
                                   + \int_0^t \biggl[
                                          \beta I(t)\norminf{h_I(t)\e^{2t}}
                                           + (1 - \delta)\norminf{h_E(t)\e^{2t}}
                                           + \norminf{L_+(\ft f_E)(t)\e^{2t}}
                                              \biggr]\, \de s\\
\norminf{h_I(\xi,t)\e^{2t}}  &\le \norminf{h_S(0)}
                                   + \int_0^t \biggl[
                                           \delta \norminf{h_E(t)\e^{2t}}
                                           + (1 - \gamma)\norminf{h_I(t)\e^{2t}}
                                           + \norminf{L_+(\ft f_I)(t)\e^{2t}}
                                              \biggr]\, \de s.
\end{aligned}
\right.
%\label{eq:seiscompetenzaFourier5}
\end{equation*}
Thanks to mass conservation we can also write
\[
\begin{aligned}
\sum_{\mathclap{H \in \ins{S,E,I}}} \norminf{L_+(\ft f_H)(t)}
                                    &= \sum_{\mathclap{H,J \in \ins{S,E,I}}}
                                       \lambda_{CJ}^2 H(t)\norminf{h_J(t)}
                                       + \sum_{\mathclap{H,J \in \ins{S,E,I}}}
                                       \expvalue{A^2_{HJ}}J(t)\norminf{h_H(t)}\\
                                    &= \sum_{\mathclap{H,J \in \ins{S,E,I}}}
                                       \bigl[\lambda_{CJ}^2 + 
                                       \expvalue{A^2_{HJ}}
                                       \bigr]H(t)\norminf{h_J(t)}\\
                                    &= \max_{H,J \in \ins{S, E, I, R}}
                                       \bigl[\lambda_{CJ}^2 + 
                                       \expvalue{A^2_{HJ}}
                                       \bigr]\sum_{\mathclap{J \in 
                                       \ins{S,E,I}}}
                                             \norminf{h_J(t)}.
\end{aligned}
\]
Now we can take advantage of condition^^>\eqref{eq:nucondition} to get
\[
\sum_{\mathclap{H \in \ins{S,E,I}}} \norminf{L_+(\ft f_H)(t)\e^{2t}}
                                    \le
                                    \nu\sum_{\mathclap{J \in \ins{S,E,I}}}
                                        \norminf{h_J(t)\e^{2t}},
\]
with $\nu < 1$. If we sum the equations of the system we therefore have
\[
\sum_{\mathclap{J \in \ins{S,E,I}}} \norminf{h_J(t)\e^{2t}} 
                                    \le
                                    \sum_{\mathclap{J \in \ins{S,E,I}}}
                                    \norminf{h_J(0)}
                                    + \int_0^t
                                    \sum_{\mathclap{\quad J \in \ins{S,E,I}}}
                                    (1 + \nu)\norminf{h_J(s)\e^{2s}}.
\]
Now Gronwall's Lemma implies
\[
\sum_{\mathclap{J \in \ins{S,E,I}}} \norminf{h_J(t)\e^{2t}}
                                    \le
                                    \norminf{h_J(0)}\e^{(1 + \nu)t}
\]
which is equivalent to
\[
\sum_{\mathclap{J \in \ins{S,E,I}}} \norminf{h_J(t)}
                                    \le
                                    \norminf{h_J(0)}\e^{-(1 - \nu)t}.
\]
Recalling the relation^^>\eqref{eq:relationd2hJ} we obtain the thesis of 
Theorem^^>\ref{teo:fouriermetric}.

\section{Structure-preserving methods}\label{appendix:B}
Here we provide some details on the structure-preserving numerical 
scheme^^>\cite{sscp} for the general class of nonlinear Fokker-Planck equations 
of the form
\begin{equation}\label{eq:sscpgeneraldbis}
\left\lbrace
\begin{aligned}
\pd{g(x,t)}{t} &= \nabla_x \cdot \bigl[\B[g](x,t)g(x,t) + \nabla_x (D(x) g(x,t))
                                  \bigr],\\
         g(x,0) &= g_0(x),
\end{aligned}
\right.
\end{equation}
where $t \ge 0$, $x \in X \contenutoin \R^d$, $d \ge 1$ and $g(x,t) \ge 0$ is 
the unknown distribution function. As mentioned above, $\B[g]$ is a bounded 
aggregation operator and $D(\cdot)$ models diffusion.

The scheme^^>\cite{sscp} follows the work of Chang and 
Cooper^^>\cite{CHANG19701} to construct a numerical method which can preserve 
features of the solution such as its large time behavior. 

If we examine system^^>\eqref{eq:fokkerplanck4}--\eqref{eq:fokkerplanck7}, we 
notice it has a structure like the following
\begin{equation}\label{eq:sscpsplitting}
\pd{\bm f(x,t)}t = \pd{\F[\bm f](x,t)}{x} + \E(\bm f(x,t)),
\end{equation}
where $\bm f(x,t) = (f_S(x,t), f_E(x,t), f_I(x,t), f_R(x,t))^T$, $\E(\bm f(x,t))$ is a 
vector accounting for dissemination dynamics
\[
\E(\bm f(x,t)) =
\begin{pmatrix}
-K(x,t)f_S(x,t) + (1-\alpha(x))\gamma(x)f_I(x,t)\\
 K(x,t)f_S(x,t) -\delta(x)f_E(x,t)\\
\delta(x)(1 - \eta(x))f_E(x,t) - \gamma(x)f_I(x,t)\\
\delta(x)\eta(x)f_E(x,t) + \alpha(x)\gamma(x)f_I(x,t),
\end{pmatrix}
\]
and $\F[\bm f](x,t)$ is the Fokker-Planck component
\[
\F[\bm f](x,t) = 
\biggl(
\pd{}{x}[(x\lambda_J -\overline{m}(t) - 4\lambda_{BJ}m_B)f_J(x,t)] + 
\frac\sigma2 \pd{^2}{x^2}(x^2f_J(x,t))
\biggr)^T_{J \in \ins{S,E,I,R}}.
\]
Here we recognize that the $J$-th entry of $\F[\bm f](x,t)$ is precisely the 
right side of equation^^>\eqref{eq:sscpgeneraldbis} in dimension $d=1$, with 
the choices
\[
\B[f](x,t) = \lambda_J x - \overline{m_x}(t) - 4\lambda_{BJ}m_B
\]
when $\alpha>0$ and $D(x) = \sigma/2 x^2$. Hence, if we consider 
system^^>\eqref{eq:fokkerplanck4}--\eqref{eq:fokkerplanck7} in the form^^>\eqref{eq:sscpsplitting} we can apply the structure-preserving 
numerical scheme^^>\cite{sscp} to it: if we consider a spatially-uniform 
grid^^>$x_i \in X$, such that $x_{i+1} - x_i = \Delta x$, and denoting $x_{i\pm 
2} = x_i \pm \Delta x/2$, we have that the discretization of the $J$-th 
component of^^>\eqref{eq:sscpsplitting} can be obtained by^^>\cite{sscp}
\begin{align}
\der{f_{J,i}(t)}t &= \frac{\F_{i+1/2}(t) - \F_{i-1/2}(t)}{\Delta x} + \E_{J,i}(t),\label{eq:sscpflux}
\end{align}
where
\[
 \F_{i+1/2} =  \C_{i+1/2} \tilde f_{i+1/2} + 
D_{i+1/2}\frac{f_{i+1}-f_i}{\delta x},
\]
having defined
\[
 \C_{i+1/2} = \frac{D_{i+1/2}}{\Delta x}\int_{x_i}^{x_{i+1}}
             \frac{\B[f](x,t) + \partial_x D(x)}{D(x)},
\]
and
\[
\tilde f_{i+1/2} = (1-\delta_{i+1/2})f_{i+1} + \delta_{i+1/2} f_i,
\]
where
\[
\delta_{i+1/2} = \frac1{\lambda_{i+1/2}} + \frac1{1-\exp(\lambda_{i+1/2})},
\]
and finally
\[
\lambda_{i+1/2} = \int_{x_i}^{x_{i+1}} \frac{\B[f](x,t) + \partial_x D(x)}
                                            {D(x)}\, \de x.
\]

For what concerns integration with respect to 
the competence level was performed using a Gauss-Legendre quadrature with 6 
points. Notice also that we need to truncate the domain of computation for
$x>0$: following^^>\cite{sscp} we  
imposed on the last grid point $x_{N+1}$ the \emph{quasi-stationary condition}^^>\cite{sscp}
\[
\frac{f_{N+1}(t)}{f_{N}(t)} = \exp\ins*{\int_{x_N}^{x_{N + 1}}
                                      \frac{\B[f](x,t) + \partial_x D(x)}
                                           {D(x)}\, \de x}
\]
Time integration of^^>\eqref{eq:sscpflux} was performed using a semi-implicit 
scheme
\[
f_i^{n+1} = f_i^n + \Delta t
            \frac{\hat \F_{i + 1/2}^{n+1} - \hat \F_{i - 1/2}^{n+1}}
            {\Delta x}+\Delta t \,\E_{J,i}^n,
\]
where
\[
\hat \F_{i+1/2}^{n+1} = \tilde \C_{i+1/2}^n
                        \Bigl[(1 - \delta_{i+1/2}^n)f_{i+1}^{n+1}
                              + \delta_{i+1/2}^n f{i+1}^{n+1}\Bigr]
                        + D_{i+1/2}\frac{f_{i+1}^{n+1} - f_i^{n+1}}{\Delta x},
\]
which, upon choosing $\Delta t = \mathcal O(\Delta x)$, preserves the 
nonnegativity of 
the solution (see^^>\cite{sscp}).
%%-----------------------------
%%      bibliography
%%-----------------------------
\bibliographystyle{abbrv}
\bibliography{refs}

\begin{thebibliography}{10}

\bibitem{network1}
G.~Albi, L.~Pareschi, and M.~Zanella.
\newblock {Opinion dynamics over complex networks: Kinetic modelling and
  numerical methods}.
\newblock {\em Kinetic \& Related Models}, 10(1):1--32, 2017.

\bibitem{Allcott}
H.~Allcott and M.~Gentzkow.
\newblock Social media and fake news in the 2016 election.
\newblock {\em Journal of Economic Perspectives}, 31(2):211--36, May 2017.

\bibitem{Bak-Colemane2025764118}
J.~B. Bak-Coleman, M.~Alfano, W.~Barfuss, C.~T. Bergstrom, M.~A. Centeno, I.~D.
  Couzin, J.~F. Donges, M.~Galesic, A.~S. Gersick, J.~Jacquet, A.~B. Kao, R.~E.
  Moran, P.~Romanczuk, D.~I. Rubenstein, K.~J. Tombak, J.~J. Van~Bavel, and
  E.~U. Weber.
\newblock Stewardship of global collective behavior.
\newblock {\em Proceedings of the National Academy of Sciences}, 118(27), 2021.

\bibitem{BertagliaPareschi}
G.~Bertaglia and L.~Pareschi.
\newblock {Hyperbolic compartmental models for epidemic spread on networks with
  uncertain data: Application to the emergence of {Covid-19 in Italy}}.
\newblock {\em Math. Mod. Meth. Appl. Sci.}, to appear (arXiv:2105.14258),
  2021.

\bibitem{BETTENCOURT2006513}
L.~M. Bettencourt, A.~Cintrón-Arias, D.~I. Kaiser, and C.~Castillo-Chávez.
\newblock {The power of a good idea: Quantitative modeling of the spread of
  ideas from epidemiological models}.
\newblock {\em Physica A: Statistical Mechanics and its Applications},
  364:513--536, 2006.

\bibitem{rey}
S.~Billiard, M.~Derex, L.~Maisonneuve, and T.~Rey.
\newblock Convergence of knowledge in a stochastic cultural evolution model
  with population structure, social learning and credibility biases.
\newblock {\em Mathematical Models and Methods in Applied Sciences},
  30(14):2691--2723, 2020.

\bibitem{networks}
L.~Brisson, P.~Collard, M.~Collard, and E.~Stattner.
\newblock {Information Dissemination in Scale-Free Networks: Profusion Versus
  Scarcity}.
\newblock In C.~Cherifi, H.~Cherifi, M.~Karsai, and M.~Musolesi, editors, {\em
  Complex Networks {\&} Their Applications VI}, pages 909--920, Cham, 2018.
  Springer International Publishing.

\bibitem{fninfoth}
D.~Brody and D.~Meier.
\newblock How to model fake news.
\newblock {\em ArXiv:1809.0096409}, 2018.

\bibitem{CHANG19701}
J.~Chang and G.~Cooper.
\newblock A practical difference scheme for {Fokker-Planck} equations.
\newblock {\em Journal of Computational Physics}, 6(1):1--16, 1970.

\bibitem{Cheng}
J.-J. Cheng, Y.~Liu, B.~Shen, and W.-G. Yuan.
\newblock An epidemic model of rumor diffusion in online social networks.
\newblock {\em The European Physical Journal B}, 86, 01 2013.

\bibitem{Conroy20151}
N.~Conroy, V.~Rubin, and Y.~Chen.
\newblock {Automatic deception detection: Methods for finding fake news}.
\newblock {\em Proceedings of the Association for Information Science and
  Technology}, 52(1):1--4, 2015.

\bibitem{fake1}
R.~P. Curiel and H.~G. Ram\'irez.
\newblock {Vaccination strategies against COVID‐19 and the diffusion of
  anti‐vaccination views}.
\newblock {\em Nature Scientific Reports}, 11:6626, 2021.

\bibitem{Daley19641118}
D.~Daley and D.~Kendall.
\newblock Epidemics and rumours.
\newblock {\em Nature}, 204(4963):1118, 1964.

\bibitem{daley_gani_1999}
D.~J. Daley and J.~Gani.
\newblock {\em {Epidemic Modelling: An Introduction}}.
\newblock Cambridge Studies in Mathematical Biology. Cambridge University
  Press, 1999.

\bibitem{PhysRevE.102.022303}
G.~Dimarco, L.~Pareschi, G.~Toscani, and M.~Zanella.
\newblock Wealth distribution under the spread of infectious diseases.
\newblock {\em Phys. Rev. E}, 102:022303, Aug 2020.

\bibitem{PhysRevLett.89.108701}
V.~M. Eguiluz and K.~Klemm.
\newblock Epidemic threshold in structured scale-free networks.
\newblock {\em Phys. Rev. Lett.}, 89:108701, Aug 2002.

\bibitem{Gelfert201884}
A.~Gelfert.
\newblock {Fake news: A definition}.
\newblock {\em Informal Logic}, 38(1):84--117, 2018.

\bibitem{Hethcote2000TheMO}
H.~Hethcote.
\newblock The mathematics of infectious diseases.
\newblock {\em SIAM Rev.}, 42:599--653, 2000.

\bibitem{rumorsontwitter}
F.~Jin, E.~Dougherty, P.~Saraf, Y.~Cao, and N.~Ramakrishnan.
\newblock Epidemiological modeling of news and rumors on twitter.
\newblock In {\em Proceedings of the 7th Workshop on Social Network Mining and
  Analysis}, pages 1--9, 2013.

\bibitem{KermackACT}
W.~O. Kermack and {\`A}.~McKendrick.
\newblock Contributions to the mathematical theory of epidemics. {II.} ---
  {The} problem of endemicity.
\newblock {\em Proceedings of The Royal Society A: Mathematical, Physical and
  Engineering Sciences}, 138:55--83, 1932.

\bibitem{korobeinikov}
A.~Korobeinikov.
\newblock Lyapunov functions and global properties for {SEIR} and {SEIS}
  epidemic models.
\newblock {\em Mathematical Medicine and Biology: {A} {J}ournal of the IMA},
  21:75--83, 07 2004.

\bibitem{PhysRevLett.86.2909}
M.~Kuperman and G.~Abramson.
\newblock Small world effect in an epidemiological model.
\newblock {\em Phys. Rev. Lett.}, 86:2909--2912, Mar 2001.

\bibitem{LEEDER2019100967}
C.~Leeder.
\newblock How college students evaluate and share \lq\lq fake new\rq\rq\
  stories.
\newblock {\em Library \& Information Science Research}, 41(3):100967, 2019.

\bibitem{fake2}
S.~Loomba, A.~de~Figueiredo, and S.~P. et~al.
\newblock {Measuring the impact of COVID-19 vaccine misinformation on
  vaccination intent in the UK and USA}.
\newblock {\em Nat. Hum. Behav.}, 5:337--348, 2021.

\bibitem{maleki2021using}
M.~Maleki, E.~Mead, M.~Arani, and N.~Agarwal.
\newblock {Using an Epidemiological Model to Study the Spread of Misinformation
  during the Black Lives Matter Movement}.
\newblock In {\em International Conference on Fake News, Social Media
  Manipulation and Misinformation (ICFNSMMM)}, 2021.

\bibitem{PIAAC19}
OECD.
\newblock {\em {Skills Matter: Additional Results from the Survey of Adult
  Skills}}.
\newblock OECD Skills Studies, OECD Publishing, Paris, 2019.

\bibitem{intermultiagent}
L.~Pareschi and G.~Toscani.
\newblock {\em Interacting multiagent systems. Kinetic equations and Monte
  Carlo methods}.
\newblock Oxford University Press, 2013.

\bibitem{wealthPareschiToscani}
L.~Pareschi and G.~Toscani.
\newblock Wealth distribution and collective knowledge. a {Boltzmann} approach.
\newblock {\em Philosophical transactions. Series A, Mathematical, physical,
  and engineering sciences}, 372, 01 2014.

\bibitem{PARESCHI2017201}
L.~Pareschi, P.~Vellucci, and M.~Zanella.
\newblock Kinetic models of collective decision-making in the presence of
  equality bias.
\newblock {\em Physica A: Statistical Mechanics and its Applications},
  467:201--217, 2017.

\bibitem{sscp}
L.~Pareschi and M.~Zanella.
\newblock Structure preserving schemes for nonlinear {Fokker–Planck}
  equations and applications.
\newblock {\em Journal of Scientific Computing}, 74:1575--1600, 03 2018.

\bibitem{PhysRevLett.86.3200}
R.~Pastor-Satorras and A.~Vespignani.
\newblock Epidemic spreading in scale-free networks.
\newblock {\em Phys. Rev. Lett.}, 86:3200--3203, Apr 2001.

\bibitem{PIQUEIRA2020123406}
J.~R. Piqueira, M.~Zilbovicius, and C.~M. Batistela.
\newblock {Daley–Kendal} models in fake-news scenario.
\newblock {\em Physica A: Statistical Mechanics and its Applications},
  548:123406, 2020.

\bibitem{Ruchansky2017797}
N.~Ruchansky, S.~Seo, and Y.~Liu.
\newblock {CSI: A hybrid deep model for fake news detection}.
\newblock In {\em International Conference on Information and Knowledge
  Management, Proceedings}, volume Part F131841, pages 797--806, 2017.

\bibitem{Shin2018278}
J.~Shin, L.~Jian, K.~Driscoll, and F.~Bar.
\newblock {The diffusion of misinformation on social media: Temporal pattern,
  message, and source}.
\newblock {\em Computers in Human Behavior}, 83:278--287, 2018.

\bibitem{IEEE}
G.~Shrivastava, P.~Kumar, R.~Ojha, P.~Srivastava, S.~Mohan, and G.~Srivastava.
\newblock Defensive modeling of fake news through online social networks.
\newblock {\em IEEE Transactions on Computational Social Systems}, PP:1--9, 08
  2020.

\bibitem{datarepository}
K.~Shu, D.~Mahudeswaran, S.~Wang, D.~Lee, and H.~Liu.
\newblock {FakeNewsNet: A Data Repository with News Content, Social Context,
  and Spatiotemporal Information for Studying Fake News on Social Media}.
\newblock {\em Big Data}, 8(3):171--188, 2020.

\bibitem{Shu2017FakeND}
K.~Shu, A.~Sliva, S.~Wang, J.~Tang, and H.~Liu.
\newblock {Fake News Detection on Social Media: A Data Mining Perspective}.
\newblock {\em ArXiv:1708.01967}, 2017.

\bibitem{trammell}
T.~I. Trammell.
\newblock {\em {Fake News Risk: Modeling Management Decisions to Combat
  Disinformation}}.
\newblock PhD thesis, Stanford University, 2020.

\bibitem{Vargo20182028}
C.~Vargo, L.~Guo, and M.~Amazeen.
\newblock {The agenda-setting power of fake news: A big data analysis of the
  online media landscape from 2014 to 2016}.
\newblock {\em New Media and Society}, 20(5):2028--2049, 2018.

\bibitem{Zhang2020}
X.~Zhang and A.~Ghorbani.
\newblock An overview of online fake news: {C}haracterization, detection, and
  discussion.
\newblock {\em Information Processing and Management}, 57(2), 2020.

\bibitem{zhao2019fake}
Z.~Zhao, J.~Zhao, Y.~Sano, O.~levy, H.~Takayasu, M.~Takayasu, D.~Li, J.~Wu, and
  S.~Havlin.
\newblock Fake news propagate differently from real news even at early stages
  of spreading.
\newblock {\em EPJ Data Sci.}, 9(7):1--14, 2020.

\end{thebibliography}
\end{document}